%% file: macs0647jd.tex
\documentclass[twocolumn]{aastex631}

\usepackage{}
\usepackage{gensymb}  
\usepackage{multirow}  

\submitjournal{ApJ}

\shorttitle{JWST MIRI of MACS0647--JD}
\shortauthors{Hsiao et al.}

\begin{document}

\title{JWST MIRI detections of H$\alpha$ and [O\,{\sc iii}] 
and direct metallicity measurement of
the $z=10.17$ lensed galaxy MACS0647$-$JD}

\correspondingauthor{Tiger Hsiao, Javier Álvarez-Márquez}
\email{tiger.hsiao@cfa.harvard.edu, jalvarez@cab.inta-csic.es}


\newcommand{\STScI}{\affiliation{Space Telescope Science Institute (STScI), 3700 San Martin Drive, Baltimore, MD 21218, USA}}

\newcommand{\JHU}{\affiliation{Center for Astrophysical Sciences, Department of Physics and Astronomy, The Johns Hopkins University, 3400 N Charles St. Baltimore, MD 21218, USA}}

\newcommand{\Stockholm}{\affiliation{Department of Astronomy, Stockholm University, Oscar Klein Centre, AlbaNova University Centre, 106 91 Stockholm, Sweden}}

\newcommand{\UMD}{\affiliation{Department of Astronomy, University of Maryland, College Park, 20742, USA}}

\newcommand{\NASAGoddard}{\affiliation{Observational Cosmology Lab, NASA Goddard Space Flight Center, Greenbelt, MD 20771, USA}}

\newcommand{\CAB}{\affiliation{Centro de Astrobiología (CAB), CSIC-INTA, Ctra. de Ajalvir km 4, Torrejón de Ardoz, E-28850, Madrid, Spain}}

\newcommand{\RIT}{\affiliation{School of Physics and Astronomy, Rochester Institute of Technology, 84 Lomb Memorial Drive, Rochester, NY 14623, USA}}

\newcommand{\ESAAURA}{\affiliation{Association of Universities for Research in Astronomy (AURA), Inc.
for the European Space Agency (ESA)}}

\newcommand{\EqualContributions}{\altaffiliation{Both lead authors contributed equally}}

\newcommand{\Uppsala}{\affiliation{Observational Astrophysics, Department of Physics and Astronomy, Uppsala University, Box 516, SE-751 20 Uppsala, Sweden}}

\newcommand{\SCAS}{\affiliation{Swedish Collegium for Advanced Study, Linneanum, Thunbergsv\"a{}gen 2, SE-752 38 Uppsala, Sweden}}

\newcommand{\Groningen}{\affiliation{Kapteyn Astronomical Institute, University of Groningen, P.O. Box 800, 9700AV Groningen, The Netherlands}}

\newcommand{\Northwestern}{\affiliation{Department of Physics and Astronomy, Northwestern University, 2145 Sheridan Road, Evanston, IL, 60208, USA}}

\newcommand{\austin}{\affiliation{Department of Astronomy, The University of Texas at Austin, Austin, TX 78712, USA}}

\newcommand{\bgu}{\affiliation{Physics Department, Ben-Gurion University of the Negev, P.O. Box 653, Be\'er-Sheva 84105, Israel}}

\newcommand{\Manchester}{\affiliation{Jodrell Bank Centre for Astrophysics, University of Manchester, Oxford Road, Manchester M13 9PL, UK}}

\newcommand{\NPPFellow}{\altaffiliation{NASA Postdoctoral Fellow}}

\newcommand{\Granada}{\affiliation{Instituto de Astrof\'isica de Andaluc\'ia, Glorieta de la Astronom\'ia s/n, 18008 Granada, Spain}}

\newcommand{\MCTI}{\affiliation{Observatório Nacional - MCTI (ON), Rua Gal. José Cristino 77, São Cristóvão, 20921-400, Rio de Janeiro, Brazil}}

\newcommand{\INAFOAS}{\affiliation{INAF--OAS, Osservatorio di Astrofisica e Scienza dello Spazio di Bologna, via Gobetti 93/3, I-40129 Bologna, Italy}}

\newcommand{\UMich}{\affiliation{Department of Astronomy, University of Michigan, 1085 S. University Ave, Ann Arbor, MI 48109, USA}}

\newcommand{\CfA}{\affiliation{Center for Astrophysics \text{\textbar} Harvard \& Smithsonian, 60 Garden Street, Cambridge, MA 02138, USA}}

\newcommand{\CIERA}{\affiliation{Center for Interdisciplinary Exploration and Research in Astrophysics (CIERA), Northwestern University, 1800 Sherman Avenue, Evanston, IL, 60201, USA.}}


\author[0000-0003-4512-8705]{Tiger Yu-Yang Hsiao} \EqualContributions \CfA \JHU \STScI 

\author[0000-0002-7093-1877]{Javier Álvarez-Márquez}\EqualContributions \CAB

\author[0000-0001-7410-7669]{Dan Coe} \STScI \JHU \ESAAURA

\author[0000-0003-2119-277X]{Alejandro Crespo Gómez} \CAB

\author[0000-0002-5258-8761]{Abdurro'uf} \JHU \STScI

\author[0000-0001-8460-1564]{Pratika Dayal} \Groningen

\author[0000-0003-2366-8858]{Rebecca L. Larson} \RIT

\author[0000-0001-8068-0891]{Arjan Bik} \Stockholm

\author[0009-0005-5448-5239]{Carmen Blanco-Prieto} \CAB  
\author[0000-0002-9090-4227]{Luis Colina} \CAB
\author[0000-0003-4528-5639]{Pablo Guillermo P\'erez-Gonz\'alez} \CAB  
\author[0000-0001-6820-0015]{Luca Costantin} \CAB  
\author[0009-0005-4109-161X]{Carlota Prieto-Jim\'enez} \CAB  

\author[0000-0002-8192-8091]{Angela Adamo} \Stockholm

\author[0000-0002-7908-9284]{Larry D. Bradley} \STScI

\author[0000-0003-1949-7638]{Christopher J. Conselice} \Manchester

\author[0000-0001-7201-5066]{Seiji Fujimoto}\altaffiliation{Hubble Fellow} \austin

\author[0000-0001-6278-032X]{Lukas J. Furtak} \bgu

\author[0000-0001-6251-4988]{Taylor A. Hutchison} \NPPFellow \NASAGoddard

\author[0000-0003-4372-2006]{Bethan L. James} \STScI \ESAAURA

\author[0000-0002-6090-2853]{Yolanda Jim\'enez-Teja}\Granada \MCTI

\author[0000-0003-1187-4240]{Intae Jung} \STScI

\author[0000-0002-5588-9156]{Vasily Kokorev} \austin

\author[0000-0003-2589-762X]{Matilde Mingozzi} \STScI

\author[0000-0002-5222-5717]{Colin Norman} \JHU \STScI

\author[0000-0003-4223-7324]{Massimo Ricotti}\UMD

\author[0000-0002-7627-6551]{Jane R.~Rigby} \NASAGoddard

\author[0000-0002-7559-0864]{Keren Sharon} \UMich 

\author[0000-0002-5057-135X]{Eros Vanzella} \INAFOAS

\author[0000-0003-1815-0114]{Brian Welch} \NASAGoddard \UMD

\author[0000-0002-9217-7051]{Xinfeng Xu} \Northwestern \CIERA

\author[0000-0003-1096-2636]{Erik Zackrisson} \Uppsala \SCAS

\author[0000-0002-0350-4488]{Adi Zitrin} \bgu





\input{newcommands}


\begin{abstract}
JWST spectroscopy has revolutionized our understanding of galaxies in the early universe. 
Covering wavelengths up to 5.3$\,{\rm \mu m}$, NIRSpec can detect rest-frame optical emission lines
H$\alpha$ out to $z = 7$ and [O\,{\sc iii}] to $z = 9.5$.
Observing these lines in more distant galaxies requires longer wavelength spectroscopy with MIRI.
Here we present MIRI MRS IFU observations 
of the lensed galaxy merger MACS0647--JD at $z = 10.165$.
With exposure times of 4.2 hours in each of two bands {(SHORT and LONG)}, we detect
H$\alpha$ at 9$\sigma$, 
[O\,{\sc iii}]$\,\lambda5008$ at 11$\sigma$, and 
[O\,{\sc iii}]$\,\lambda4960$ at 3$\sigma$.
%
Combined with previously reported NIRSpec spectroscopy that yields seven emission lines including the auroral line [O\,{\sc iii}]$\,\lambda4363$,
we present the first direct metallicity measurement of a $z > 10$ galaxy: $12+{\rm log(O/H)}= 7.79\pm0.09$, or 0.13$^{+0.02}_{-0.03}\,Z_{\odot}$. 
This is similar to galaxies at $z \sim 4 - 9$ with direct metallicity measurements,
though higher than expected given the high specific star formation rate 
log(sSFR / yr$^{-1}$) = $-7.4 \pm 0.3$.
We further constrain
the ionization parameter ${\rm log}(U)$ = $-1.9 \pm 0.1$, 
ionizing photon production efficiency log($\xi_{\rm ion}$) = $25.3\pm0.1$,
and star formation rate $SFR = 5.0\pm0.6\,M_{\odot}/{\rm yr}$ within the past 10 Myr. 
These observations demonstrate the combined power of JWST NIRSpec and MIRI
for studying galaxies in the first 500 million years.
\end{abstract}


\section{Introduction} \label{sec:intro}

JWST was designed to study galaxies in the early universe
\citep{Gardner2023,Rigby2023},
including the first galaxies composed of long-theorized Population III stars
made of only the primordial elements hydrogen and helium formed in the Big Bang.
Such pristine stars and galaxies would have no heavier elements and zero gas-phase metallicity. 

While we have yet to find the elusive first-generation Pop III stars,
we have measured the build-up of metals in galaxies
over 13 billion of years of cosmic history.
On average, gas-phase metallicity increases over time (decreases with redshift),
increases with stellar mass \citep[the mass-metallicity relation; e.g.,][]{Tremonti2004,Nakajima2023}, 
and decreases with specific star formation rate \citep[the Fundamental Metallicity Relation; e.g.,][]{Mannucci2010,Curti2020}.

Ground-based telescopes have measured galaxy metallicities out to $z = 3.5$
based on measurements of rest-frame optical lines, 
including \OIIIw, \Hbeta, and \Halpha. The fainter auroral line \OIIIwa, when detected, 
provides the most precise metallicity measurement
based on the ``direct'' method of deriving the electron temperature $T_e$
from the line flux ratio \OIIIw\ $/$ $\lambda$4363 \citepeg{Peimbert1967,Osterbrock1989,Izotov2006,Kewley2008,Andrews2013,Sanders2024}.
Stronger \OIIIwa\ corresponds to higher temperatures $T_e$ and lower metallicity.

JWST NIRSpec spectroscopy covering 0.6--5.3 \um\ \citep{NIRSpec_Jakobsen2022,NIRSpec-MOS_Ferruit2022,Boker2023} has enabled direct metallicity measurements in galaxies out to $z \sim 9$ observed 500 Myr after the Big Bang.
Most galaxies at $z \sim 4$ -- 9 observed to date have metallicities 
ranging from 3\% to 50\% \Zsun\ (including direct method and strong line calibrations),
or \logOH\ $= 7.2$ -- 8.4
\citep[e.g.,][]{Curti2023, Nakajima2023,Boyett2023,Morishita2024}.
The most notable exception is the $z = 6.64$ strongly lensed galaxy LAP1 
with very low metallicity $<$ 0.4\% \Zsun, or \logOH\ $< 6.3$ \citep{Vanzella2023}.
%

For more distant galaxies at $z > 9.5$, \OIIIw\ redshifts beyond NIRSpec's wavelength range,
requiring the longer wavelength 5--28 \um\ coverage of MIRI \citep{Rieke2015,Wright2015,Wright2023}.
MIRI is capable of detecting strong emission lines, such as \Hbeta, \OIIIdw\, and \Halpha, in distant galaxies \citep{Alvarez2019}
as recently demonstrated for the first time at $z > 9$ 
with \Halpha\ detections in the strongly lensed galaxy MACS1149-JD1 at $z = 9.11$ \citep{Alvarez2023} 
and for the first time at $z > 10$ 
in the galaxy GHZ2/GLASS-z12, which is only weakly lensed, at $z=12.33\pm0.02$ \citep{Zavala2024}.



%


The strongly lensed galaxy MACS0647$-$JD 
was discovered in HST imaging \citep{Coe2013}.
Three lensed images JD1, JD2, and JD3 are observed with
JWST NIRCam F200W AB mag 25.0, 25.4, and 26.2, respectively,
for a total AB mag 24.2,
making it the brightest $z > 10$ galaxy known \citep{Hsiao2023a}.
The images are magnified by factors of \about8.0, 5.3, and 2.2,
based on lens modeling \citep{Zitrin2015,Chan2017,Meena2023}
and the observed flux ratios.

\citet{Hsiao2023a} reported that NIRCam imaging resolved MACS0647$-$JD as having 
two small stellar complexes A (JDA) and B (JDB)
in a possible galaxy merger.
The larger and brighter component A has an effective radius $r=70\pm24\,{\rm pc}$ and appears bluer, likely due to its young stellar population ($\sim50\,{\rm Myr}$ old) and no dust.
In contrast, the other component, JDB, is smaller, with a radius of $r = 20^{+8}_{-5}\,{\rm pc}$ and is redder, likely due to an older stellar population ($\sim$100 Myr old) and mild obscuration ($A_{V}\sim0.1\,{\rm mag}$).
The dissimilar inferred star formation histories suggest the two components formed separately and are in the process of merging.
Their projected separation is \about 400 pc (delensed).
A possible third companion C, also triply-lensed, was identified 2\arcsec\ away (\about 3$\,$kpc delensed).

JWST Cycle 1 NIRSpec prism spectroscopy
confirmed the redshift $z=10.17$ and revealed
seven emission lines, including \CIIIdw, \OIIw, \NeIIIw, \NeIIIwb, \Hdeltaw, \Hgammaw, and the auroral line \OIIIwa\ \citep{Hsiao2023b}.
Based the observed line ratios,
\cite{Hsiao2023b} estimated the metallicity using various methods
yielding values ranging from 6\% to 20\%$\,$\Zsun.
A direct metallicity measurement was not possible
due to the lack of \OIIIw.



This paper presents new MIRI Medium Resolution Spectrograph 
(MRS, \citealt{Wells2015,Argyriou2023}) 
and Mid-Infrared Imager Module (MIRIM, \citealt{Bouchet2015,Dicken+24}) 
observations of MACS0647--JD. 
The MRS spectroscopy yields detections of the \OIIIdw\ and \Halpha\ emission lines,
and the MIRI imaging shows the rest-frame optical structure of the galaxy merger.
Our measured \Halpha\ flux provides a robust measurement of the star formation rate (SFR) 
and the ionizing photon production efficiency (\xiion) when combined with existing rest-frame UV NIRCam photometry from \cite{Hsiao2023a}. 
MIRI MRS in combination with NIRSpec prism spectroscopy 
provides the first ever measurement of the gas-phase metallicity ($Z$) using the ``direct" method in a galaxy at a redshift $z > 10$. 
In addition to the metallicity, we are able to constrain key physical properties of the interstellar medium (ISM), including the electron temperature ($T_{\rm e}$) and the ionization parameter ($U$).
%

This paper is organized as follows: 
\S\ref{Sec:obs_cal} describes the JWST/MIRI observations and data reduction.
\S\ref{sec:MIRI_phot_spec} presents the MRS 1D extracted spectra, emission line detections, and flux calculations.
\S\ref{sec:result_discussion} details the main results and discusses them. 
\S\ref{sec:conclusion} summarizes and concludes the paper. 


Throughout, we use the AB magnitude system,
which relates magnitude to flux in nanoJanskys via:
AB mag $= 31.4 - 2.5 \log(F_{\rm nJy})$ \citep{Oke74,OkeGunn83}.
We adopt the {\em Planck} 2018 flat \LCDM\ cosmology \citep{Planck18_cosmo}
with $H_0 = 67.7$ km s\inv\ Mpc\inv, $\Om = 0.31$, and $\OL = 0.69$,
for which the universe is 13.8 billion years old,
and $1 \arcsec \sim 4.2$ kpc at $z = 10.165$ observed 460 million years after the Big Bang.
Where needed, we assume the \citet{Chabrier2003} initial mass function (IMF). 
We quote metallicities $Z / Z_\odot$ relative to a solar abundance \logOH\ = 8.69 \citep{Asplund2009}.
Lensing magnifications of 8.0 and 5.3 are adopted for JD1 and JD2, respectively \citep{Hsiao2023a}.
Uncertainties from lensing effect are not propagated throughout the letter.

\begin{figure*}
\centering
\includegraphics[width=\textwidth]{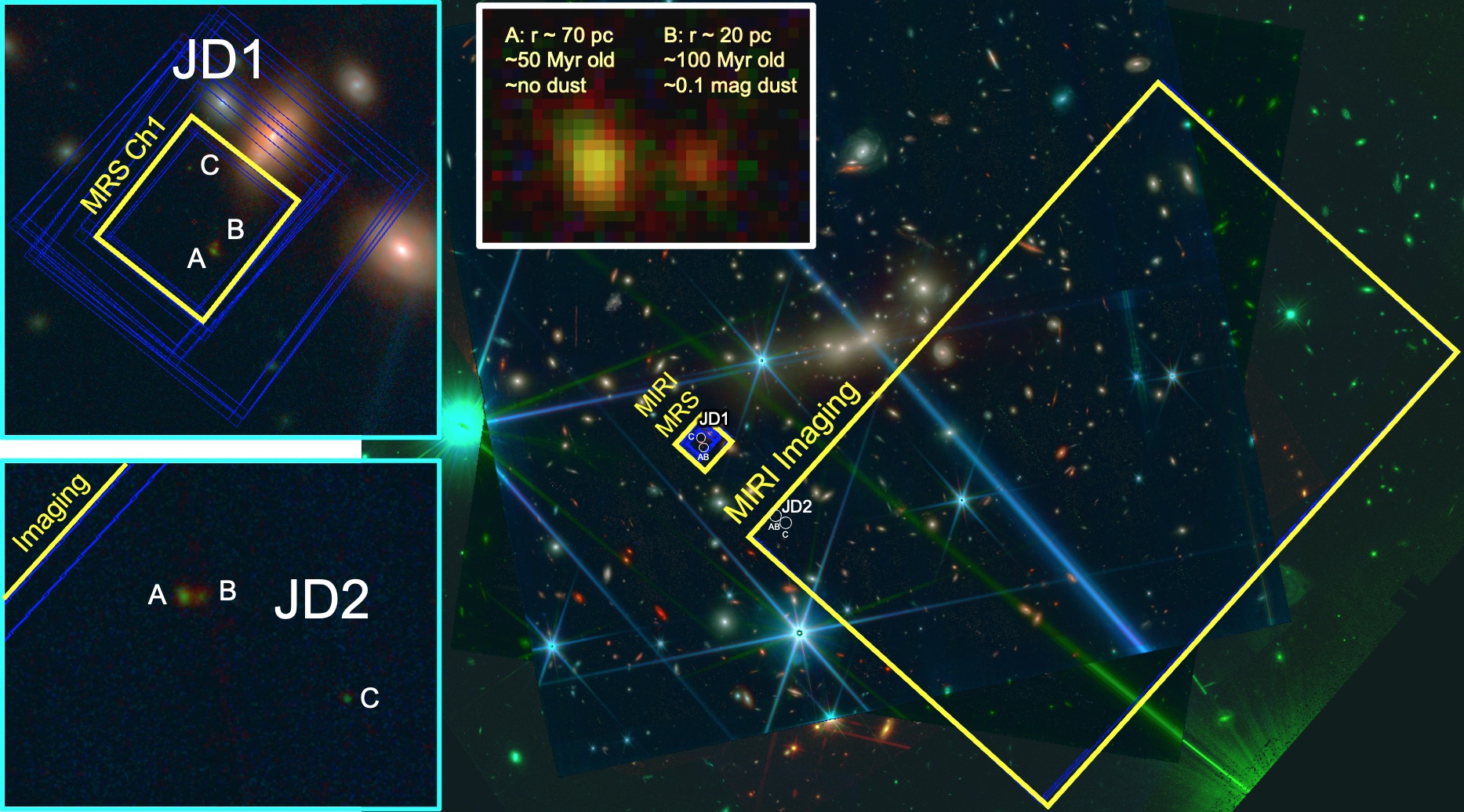}
\caption{JWST MIRI coverage of the MACS0647 field by GO 4246 (PI: Abdurro'uf) overlaid on the composite color image of JWST and HST. 
\emph{Top left:} MIRI MRS IFU spectroscopy on the brightest lensed image, MACS0647$-$JD1, including all three clumps A, B, and C.
All three components are observed in the FoV in all 4 dithers.
\emph{Bottom left:} The second brightest lensed image, MACS0647$-$JD2,
is included in the MIRI imaging field of view.
\emph{Right:} Full image of the simultaneous MIRI imaging and MRS coverage.
Note a bright 8th magnitude star lurks just below and right off the edge of the HST field of view.
Two zoom insets show the JWST NIRCam images of JD1 and JD2.
}
\label{fig:obs}
\end{figure*}

\begin{figure*}
\centering
\includegraphics[width=\textwidth]{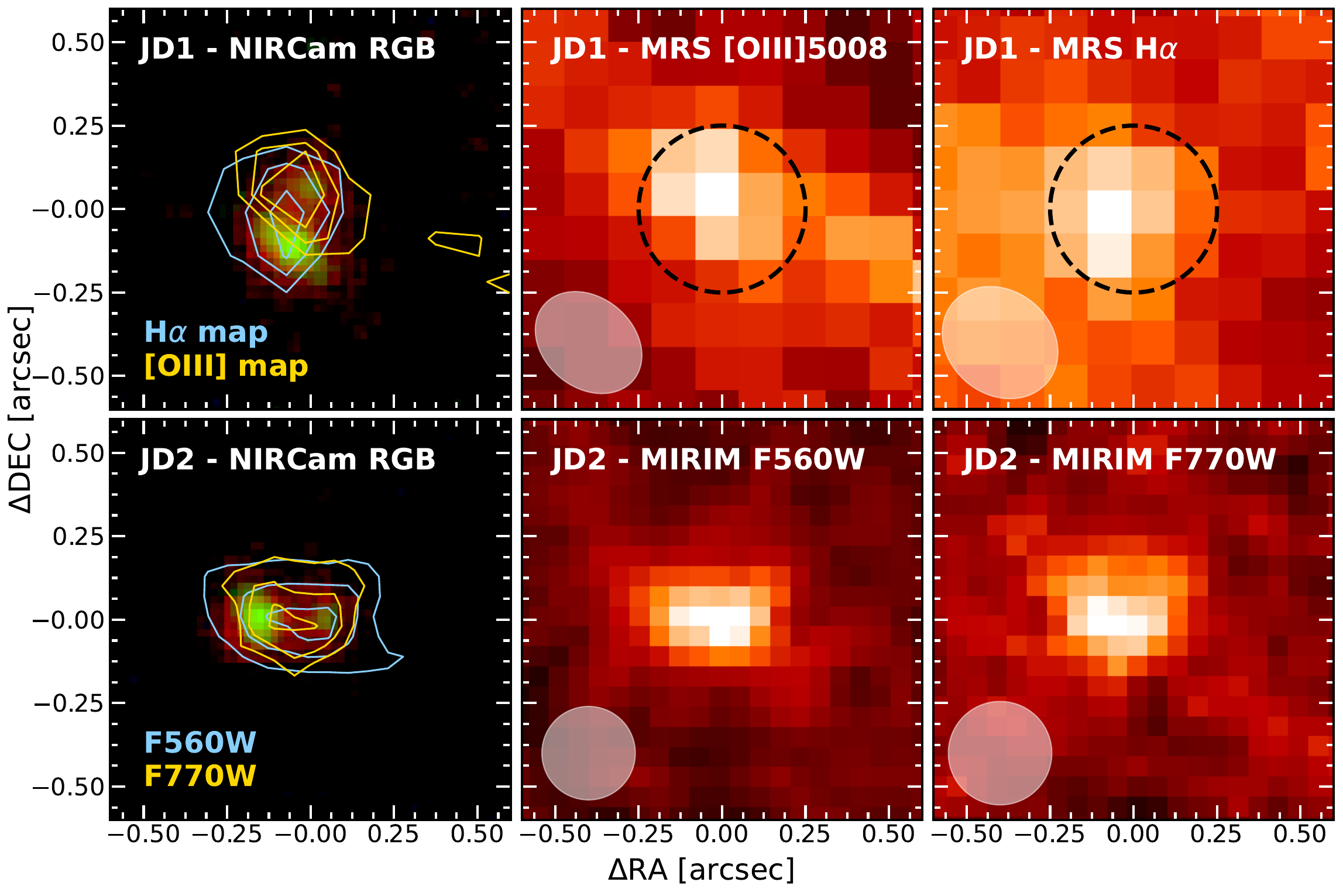}
\caption{\emph{Top row:}
JD1 NIRCam F200W image (drizzled to $0.03''$ pixels)
and MIRI MRS line maps of \OIIIw\ and \Halpha\ ($0.13''$ pixels)
integrated within $\pm 100$ \kms\ of the line peaks.
Contours from these \OIII\ (yellow) and \Halpha\ (blue) line maps are overlaid on the NIRCam image of JD1.
The dashed circle is used to measure flux shown in Figure \ref{fig:spectra}.
\emph{Bottom row:}
JD2 NIRCam F200W image and MIRIM F560W and F770W images
($0.06''$ pixels).
Contours from F560W (blue) and F770W (yellow) line maps are overlaid on the NIRCam image of JD2.
All data in each row are aligned and shown within an area $1.25'' \times 1.25''$.
The center coordinates (RA, Dec) in degrees are 
(101.98225, 70.243302) for JD1  
and
(101.97128, 70.239721) for JD2. 
The JD1 centroid is also used for the aperture.
PSF sizes are shown in the lower-left corners.
}
\label{fig:map}
\end{figure*}



\section{Observations and data reduction}
\label{Sec:obs_cal}

JWST Cycle 2 program GO 4246 (PI: Abdurro'uf) observed MACS0647-JD simultaneously
with MIRI MRS and MIRIM
on February 11, 2024 (Figure \ref{fig:obs}).
In this letter, we will only discuss the MIRI MRS data and save the imaging analysis for future work.
The data are publicly available on Mikulski Archive for Space Telescopes (MAST\footnote{\url{https://mast.stsci.edu/search/ui/\#/jwst}}{; \dataset[DOI:10.17909/re1k-jt10]{https://archive.stsci.edu/doi/resolve/resolve.html?doi=10.17909/re1k-jt10}}).
and we provide reduced MRS data products on our website.\footnote{\url{https://cosmic-spring.github.io}}

\subsection{MIRI observations}


MRS observations have a FoV of 3.2\arcsec $\times$3.7\arcsec\ in channel 1. 
Our MRS pointing includes the three components (A, B, and C) of the JD1 lensed galaxy image. 
The observations were obtained in two MRS bands, SHORT and LONG, covering the 4.90--5.74\,$\mu$m and 6.53--7.65\,$\mu$m spectral ranges for channel 1, respectively.
These spectral bands mainly target the \Hbeta\ and \OIIIww, and the \Halpha\ bright optical emission lines, respectively. 
While the MRS observed JD1, MIRIM simultaneously imaged JD2 and an adjacent region with a FoV of 74\arcsec $\times$113\arcsec\ (Figure \ref{fig:obs}).
MIRIM imaging was obtained in the F560W (5.0--6.3 \um) and F770W (6.6--8.7 \um) filters at the same time as the SHORT and LONG MRS observations, respectively. 
Total exposure times were 15194 seconds (4.2 hours) for each simultaneous MRS band + MIRIM filter. These were split among 4 dithers in a pattern optimized for extended sources. At each dither position, the 1.05-hour exposure consisted of 8 integrations of 19 groups of SLOWR1 readout.

Additional shallower observations were obtained of a nearby blank region 15\arcsec\ to the west 
to measure MRS backgrounds. With single integrations and only 2 dithers, the total exposure times for these background observations were 15 minutes in each MRS band + MIRIM filter combination. 
The MRS background observations are not used in this paper, and the backgrounds are subtracted following a different methodology described below. 
We do include all MIRIM data, including the additional background exposures, in our image reductions.
The on-source and background simultaneous MIRIM observations will be combined to generate the calibrated MIRIM images of filters F560W and F770W.

\subsection{MIRI data calibration \label{MIRI_cal_sec}}

The MRS observations are processed with version {1.13.4} of the JWST calibration pipeline and context {1215} of the Calibration Reference Data System (CRDS). We follow the standard MRS pipeline procedure \citep{Bushouse.2024}, with additional customized steps to improve the quality of the final MRS calibrated products (see \citealt{Alvarez-Marquez2023SPT,Alvarez2023} for details). 
These steps account for: 
(i) generation of a new bad and hot pixel mask, 
(ii) subtraction of the residual flux in regions of the detector that see no direct light to ensure a median count rate equal to zero and consistency in the flux calibrations, 
(iii) subtraction of 1/f noise (correlated noise in the vertical direction of the detector) using a smoothed function that additionally allows the removal of the residuals from shallow cosmic ray showers, 
(iv) generation and subtraction of a pixel-by-pixel master detector background to correct any detector artifact and fringe residuals left after the standard calibrations, 
and 
(v) sigma clipping to mask the residuals of bright cosmic ray showers. 

The compact dithers used in this program do not fully separate the emission of a point-like source in the various exposures and detector plane.
This makes it impossible to use the on-source observations to generate a pixel-by-pixel master detector background clean of any emission from JD1. 
The dedicated background observations of this program have exposure times that are too short to perform a pixel-by-pixel detector background subtraction; they would inject significant noise into the on-source observations and reduce the SNR of the final calibrated cube. 

So for step (iv) in this work, we follow a different approach than presented in the previous cited works. 
We have compiled MRS observations that are similar to MACS0647--JD: 
B14--65666, 
MACS1149--JD1 \citep{Alvarez2023}, 
and GNz11 
from the MIRI European Consortium GTO programs (IDs 1284 \& 1262) 
and the GO2 program (ID 2926).
We have calibrated these observations following the same procedure as MACS0647--JD1. 
The B14--65666 and GNz11 observations are used to generate the pixel-by-pixel master detector background of band SHORT, and MACS1149--JD1 observations are used for band LONG. 
These master backgrounds have been generated using a sigma-clipping median of all dithers, following \cite{Alvarez-Marquez2023SPT,Alvarez2023}. 
We note these corrections do not significantly affect the background level, so any systematic differences in backgrounds between the datasets won't significantly affect the measured line fluxes.

The final 1SHORT and 1LONG 3D spectral cubes have a spatial and spectral sampling of 0.13"\,$\times$\,0.13"\,$\times$\,0.8\,nm \citep{Law2023}, and a resolving power $R = \Delta \lambda / \lambda$
of about 3500 (85\,km\,s$^{-1}$, \citealt{Labiano2021, Jones2023}).  


The MIRIM observations are calibrated with version 1.13.4 of the JWST pipeline and context 1210 of the CRDS. 
In addition to the general procedure, our data calibration process includes additional steps to correct for striping artifacts and background gradients (see \citealt{Alvarez-Marquez2023SPT,Perez-Gonzalez+24} for details). 
All of the MIRIM observations cover JD2, including both MRS on-source and background.
Combining all MIRIM exposures,
we generate final dithered F560W and F770W images with pixel sizes of 0.06$\arcsec$. 
We have generated effective point spread function (PSF) combining four stars within the MIRIM FoV. These PSFs present full width half maximum (FWHM) of 0.28$\arcsec$ and 0.31$\arcsec$, which correspond to 33\% and 15\% increase in the MIRI native spatial resolution due to the compact dithers pattern used in the MRS-MIRIM simultaneous observations \citep{Libralato+24}.

\begin{figure*}
\centering
\includegraphics[width=0.49\textwidth]{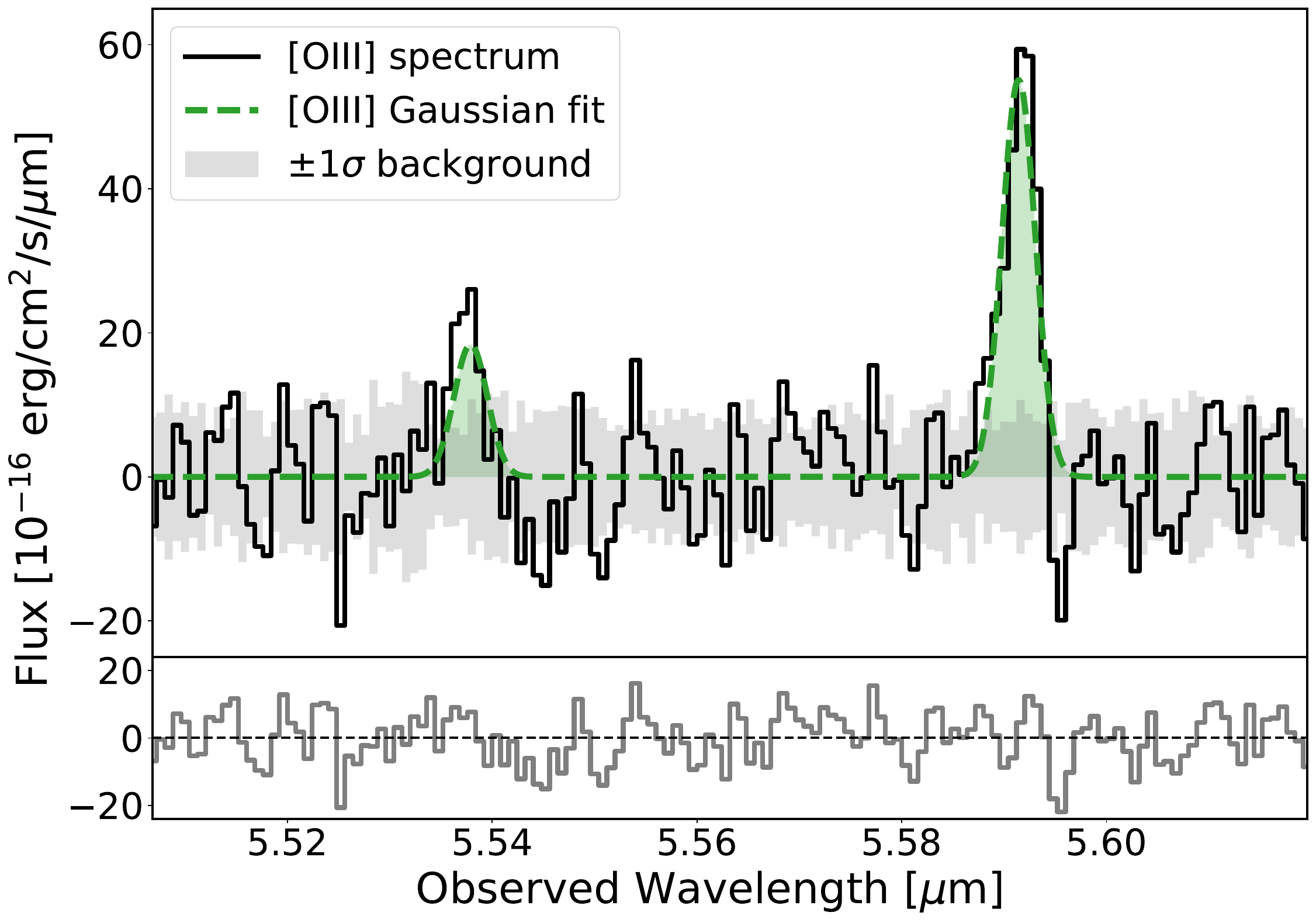}
\includegraphics[width=0.49\textwidth]{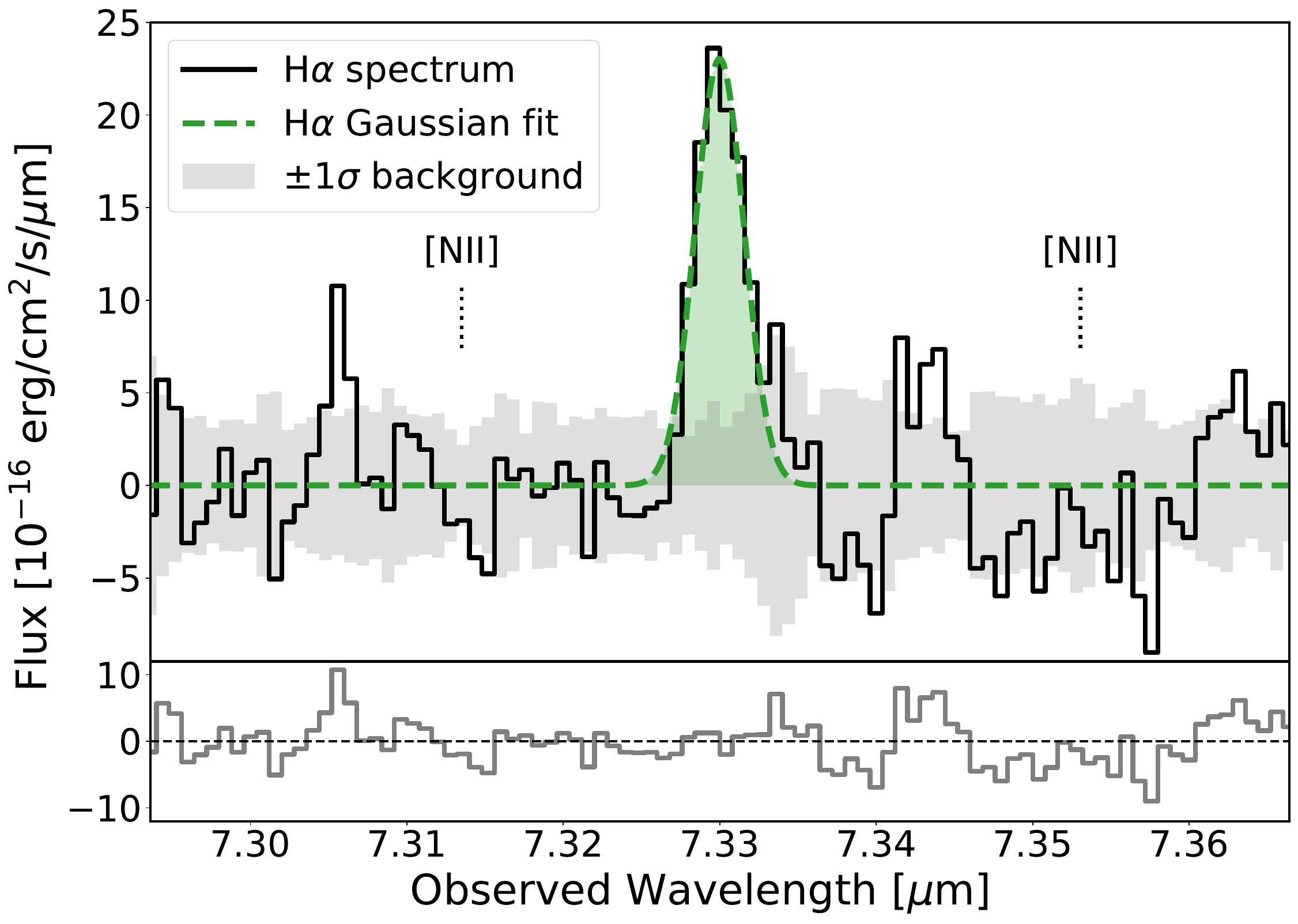}
\caption{
MIRI MRS spectra of \OIIIdw\ and \Halpha\ observed at $z = 10.165$.
\emph{Left:} \OIIIw\ and \OIIIwc\ spectra in black measured within the circular aperture shown in Figure \ref{fig:map}.
Uncertainties are shown in shaded gray.
A single Gaussian (green) is well fit to the \OIIIw\ line;
that same fit is shown for the \OIIIwc\ line after reducing the flux by a factor of 2.98.
Fit residuals are shown in the bottom plot.
\emph{Right:} 
\Halpha\ spectrum fit to a single Gaussian.
Vertical dashed lines show the wavelengths of the weaker undetected lines \NIIww\ that are well separated in wavelength and thus do not contribute to \Halphaw.
}
\label{fig:spectra}
\end{figure*}

Finally, we correct the astrometry in the MIRIM and MRS observations. MIRIM images are aligned by measuring the centroid of two field stars with available GAIA DR3 \citep{Gaia2022} coordinates. As the MIRIM and MRS have been observed simultaneously, the offset correction for the MIRIM images is also implemented in the MRS 3D spectral cubes. The final uncertainty in the data-set alignment is less than a pixel in the MIRIM drizzle images (i.e., $<$\,60\,mas). Figure \ref{fig:map} shows the MRS \OIIIw\ and \Halpha\ line maps of JD1 and the MIRIM F560W and F770W images of JD2. 
The line maps are generated by integrating the MRS cubes in the velocity range $-100$\,$<$\,$v$\,[km\,s$^{-1}$]\,$<$\,100, taking as a reference the \OIIIw\ and \Halpha\ emission line peaks derived by Gaussian line fitting (see Figure \ref{fig:spectra} and \S\ref{sec:MIRI_phot_spec}). 

\subsection{NIRSpec and NIRCam ancillary data \label{sec:NIRSpec_NIRCam}}

This letter also makes use of the NIRCam and NIRSpec observations of JWST Cycle 1 program GO 1433 (PI Coe).
NIRCam imaging was obtained in 7 filters, F115W, F150W, F200W, F277W, F365W, F444W, and F480M spanning 1--5$\,\mu$m.
The NIRCam images were reduced by the STScI JWST pipeline and \grizli\ \citep{grizli}.
{ Photometry was reported in \citet{Hsiao2023a} and updated in \citet{Hsiao2023b}.
We use measurements from the latter,
including F200W $368\pm 5\,{\rm nJy}$ and F444W $317\pm 8\,{\rm nJy}$
within a $r = 0.25\arcsec$ aperture by \grizli, which also includes aperture corrections to total flux using \textsc{SourceExtractor} \texttt{MAG\_AUTO}.
We adopt these as total flux measurements as they are more inclusive (capturing more flux) than \textsc{GALFIT}\footnote{\url{https://users.obs.carnegiescience.edu/peng/work/galfit/galfit.html}} \citep{Galfit2002} modeling both clumps A and B, where we obtained similar results (within $4\%$ for F444W) with an independent analysis using an $r = 0.6\arcsec$ and PSF-based aperture corrections.
We also modeled components A and B as Sersic profiles using \textsc{GALFIT}, 
$\sim10\%$ lower fluxes likely because they miss additional fainter components or diffuse emission.
We use the F444W measurement in \S\ref{sec:normalization} to
normalize the NIRSpec prism emission line fluxes.
}

NIRSpec multi-object spectroscopy (MOS) was performed in two observations (Obs 21 and Obs 23) using the microshutter assembly (MSA) with the low-resolution $R \sim 30-300$ prism spanning 0.6 -- 5.3$\,$\um\ \citep{Hsiao2023b}.
Obs 23 was executed with the standard 3-slitlet nods while Obs 21 was performed with single-slitlet mode.
Both Obs 21 and Obs 23 observed JD1 and JD2.
However, only Obs 21 JD2 targeted the center of MACS0647$-$JD, covering A+B (see Figure 1 in \citealt{Hsiao2023b}), while other observations mainly targeted JDA (including JD1A and JD2A).
In this paper, we use emission line fluxes measured from the stacked spectrum \citep{Hsiao2023b}
as well as individually in Obs 21 JD2 as measured by \citet{Abdurrouf2024} using \textsc{piXedfit} code \citep{Abdurrouf2021,Abdurrouf2022}.
Ultimately, we find this choice does not significantly affect the results.

\section{Spectroscopic measurements}
\label{sec:MIRI_phot_spec}

\subsection{MIRI}

MRS spectroscopy has detected and spectrally resolved the rest-frame optical emission lines \OIIIdw\ and \Halpha\ of JD1 (see Figures \ref{fig:map} and \ref{fig:spectra}).
JD1 rest-frame optical fluxes detected with MRS are spatially coincident with the A and B clumps identified in the NIRCam imaging covering the rest-frame UV emission (see Figure \ref{fig:map}).

Clumps A and B of JD1 are not spatially resolved in the MRS 1SHORT and 1LONG observations, as their separation (\about0.2$\arcsec$, \citealt{Hsiao2023a}) is smaller than the PSF FWHM of the MRS channel 1 ($\sim$\,0.3$\arcsec\times$0.4$\arcsec$, \citealt{Argyriou2023}). 
Thus in our analysis for this paper, we treat JD1 as an individual unresolved source. 
A small offset ($\sim$0.1$\arcsec$) between \OIIIw\ and \Halpha\ is observed in the line maps presented in Figure \ref{fig:map}. This offset could be interpreted as the combination of the low signal-to-noise ratio of the line maps, and the uncertainties derived from the astrometry correction ($<$\,60\,mas) together with the MRS geometry distortion and wheel positioning repeatability ($\sim$\,40-50\,mas; \citealt{Patapis2024}). The fainter companion clump C is within the MRS FoV but not detected in \OIIIw\ nor \Halpha.

\subsubsection{\OIIIdw\ and \Halpha\ spectra and fluxes}


We extract the integrated \OIIIdw\ and \Halpha\ 1D spectra of JD1 using a circular aperture of 0.25$\arcsec$ radius in both MRS channels (see Figure~\ref{fig:map}). We also extract twelve background 1D spectra using the same sized aperture from random positions in the MRS FoV clean of source emission. We combine the twelve background spectra to generate the 1D median and standard deviation of the local background of each channel. The median background is compatible with zero in both cases, and the standard deviation is used as a 1$\sigma$ error.  {The \OIIIw\ and \Halpha\ extracted spectra, together with the 1$\sigma$ error, are corrected for aperture losses assuming that the JD1 emission is spatially unresolved. We use the MRS PSF model of channel 1 (\citealt{Argyriou2023}; Patapis in prep.). The fraction of flux for the selected aperture is 65\,\% and 59\,\% at \OIIIw\ and \Halpha\ observed wavelengths, respectively. These factors are implemented to derive the total spectra of \OIIIdw\ and \Halpha, which are shown in Figure~\ref{fig:spectra}.}

We model the \OIIIw\ and \Halpha\ emission line spectra with Gaussian functions 
plus a second-order polynomial to fit any residual background gradient (see Figure~\ref{fig:spectra}). 
The uncertainties on the derived emission line parameters, such as FWHM, flux, and redshift, are estimated by bootstrapping the spectra accordingly with the noise. The noise of the spectrum is measured as the RMS of the continuum surrounding the emission line. This noise is used to generate 1000 new spectra with random Gaussian noise added to the original spectrum before the lines are fit again. The final uncertainty is the standard deviation of all the individual measurements. In Table \ref{tab:lines}, we present \OIIIw\ and \Halpha\ line fluxes and Gaussian parameters based on these fits. The \OIIIdw\ emission lines have been fitted together assuming a flux ratio of 2.98, the same FWHM and redshift. 



We detect \OIIIwc\ and \OIIIw\ with signal-to-noise ratios (SNRs) 3.6 and 10.7, and fluxes ($75 \pm 21$)\e{-19} and ($226 \pm 21$)\e{-19} \cgsfluxunits, respectively. \Halpha\ is detected with SNR 9.0 and flux ($90 \pm 10$)\e{-19} \cgsfluxunits.  
The redshift calculated from the location of the peak of the Gaussian fit is 10.1644$\pm$0.0005 and 10.1659$\pm$0.0004 for \OIIIw\ and \Halpha\, respectively. 
The \OIIIw\ and \Halpha\ lines have intrinsic FWHM of $191\pm19$ \kms\ and $127\pm19$ \kms, respectively, after correcting for the instrumental line broadening \citep{Labiano2021}, and an velocity offset of 39$\pm$20 \kms. 
The redshift offset and FWHM difference between the two lines 
are due to the asymmetry of the \OIIIw\ emission line, which blueshifts the peak location and broadens the profile of the Gaussian fit. 
The \OIIIw\ line could alternatively be modeled as a sum of two Gaussians, 
with the main component having the same redshift and FWHM as \Halpha. 
Analysis of the \OIIIw\ line profile asymmetry is outside the scope of this paper and will be presented in \'Alvarez-M\'arquez in prep.

Fainter emission lines covered by the 1SHORT and 1LONG wavelength ranges were not detected. 
Assuming Case B recombination, we expect a flux ratio \Halpha\ $/$ \Hbeta\ = 2.86
and an \Hbeta\ flux $\sim 3$\e{-18} \cgsfluxunits.
Thus we should only expect SNR \about\ 1.5 in MRS 1SHORT
(which is slightly less sensitive than 1LONG)
given the 1\sig\ flux uncertainties $\sim 2$\e{-18} \cgsfluxunits\ at 
the \Hbeta\ rest-frame wavelength 4863\AA\ near \OIIIww.
Still fainter lines eluding detection include \NIIww\ and \SIIww\ with 
1\sig\ uncertainty \about \tentotheminus{18} \cgsfluxunits,
or 3\sig\ upper limits \about 3\e{-18} \cgsfluxunits.




\input{lines}  

\subsection{NIRSpec spectroscopy normalized to MIRI MRS}
\label{sec:normalization}

We assume our MIRI MRS line flux measurements are the total values for JD1 A+B,
given the complete coverage with the IFU and the aperture loss corrections we applied to our 0.25\arcsec\ aperture.
We cannot say the same for our NIRSpec MSA line fluxes
since the slits do not fully cover JD1 A+B, and our slit loss corrections are uncertain.
The JWST pipeline corrects for slit losses assuming either a point source or uniform illumination;
we choose the point source option.
More robust slit loss estimates would require modeling the two extended sources, as in \cite{deGraaff2023}.

{In \cite{Hsiao2023b}, we measured NIRCam photometry within rectangular apertures matched to the NIRSpec MSA slits.
However, this technique assumes uniform transmission through the slits, not accounting for slit losses.}
Here, we integrate the NIRSpec prism spectra over the F444W filter, 
measuring 105 nJy for the stacked spectrum (with JD1 magnification $\mu = 8$),
yielding a flux loss factor of 3.0
{ (compared to $317 \pm 8$ nJy measured in the NIRCam image; \S\ref{sec:NIRSpec_NIRCam})}.
Thus we multiply our NIRSpec line flux measurements by a factor of 3 to correct for slit losses
and apply other factors as needed to correct for magnification 
{ (JD1 $\mu = 8.0$ and JD2 $\mu = 5.3$).}
We use updated line flux measurements presented in \citet{Abdurrouf2024}
that are similar to those presented in \cite{Hsiao2023b} for the stacked spectrum.


Based on this normalization, we estimate \Halpha\ $/$ \Hgamma\ $= 5.5 \pm 0.7$.
We adopt this normalization as fiducial in our analysis.
This is consistent with the expected ratio
\Halpha\ / \Hgamma\ $= 6.11$ (5.79) 
for Case B recombination at $T=10000\,$K ($20000\,$K) {assuming no dust}
\citep[e.g.,][]{Dopita2003,Groves2012}.
We also consider an alternative analysis fixing the line ratio to
\Halpha\ $/$ \Hgamma\ $= 6.11$ (assuming no dust).


In addition, we perform the same procedure with the individual NIRSpec prism observation
(Obs 21 JD2) that is best centered on A+B (though not fully including both),
perhaps better corresponding to the centroid of the line emission
combined from both clumps A and B 
(see \citealt{Hsiao2023b} and \citealt{Abdurrouf2024}).
This observation may be more representative of MACS0647--JD as a whole.
Our analyses of the NIRSpec and MIRI data assume all the emission comes from a single source 
since we do not spatially resolve A and B with these observations.
\input{properties2}

\section{Result and Discussion} \label{sec:result_discussion}






\subsection{SFR and \xiion}
\label{sec:Halpha}


Given this nearly unprecedented measurement of \Halpha\ at $z > 10$
\citep[see also][]{Zavala2024},
we can directly measure the recent star formation rate (SFR)
and ionizing photon production efficiency \xiion\ of MACS0647--JD.

Based on the total measured JD1 H$\alpha$ line flux 
$(9\pm1)$\e{-18} \cgsfluxunits,
and correcting for magnification $\mu \approx 8$,
we estimate an intrinsic luminosity $L_{\rm H\alpha}=(1.6\pm0.2)\times10^{42}\,{\rm erg\,s^{-1}}$. 
Adopting an SFR conversion factor of 3.2 $\times$ \tentotheminus{42} 
suitable for high-redshift galaxies {with a lower metallicity of $0.28\,Z_{\odot}$} \citep{Reddy2018},
we estimate ${\rm SFR}=5.0\pm0.6\,M_{\odot}{\rm yr^{-1}}$, assuming zero escape fraction, no dust attenuation, and a \citet{Chabrier2003} IMF.
%
This is consistent with previous estimates
based on JWST NIRCam SED fitting \citep[$\sim4-10\,M_{*}/{\rm yr}$][]{Hsiao2023a,Hsiao2023b}.
We note that it is higher than the estimates of ${\rm SFR=1.4\pm0.2}$ from NIRSpec prism spectroscopy including \Hgamma\ \citep{Hsiao2023b}, since NIRSpec observations mainly targeted A and it is not corrected for flux losses (see also \S\ref{sec:normalization}).

\begin{figure}
\centering
\includegraphics[width=\columnwidth]{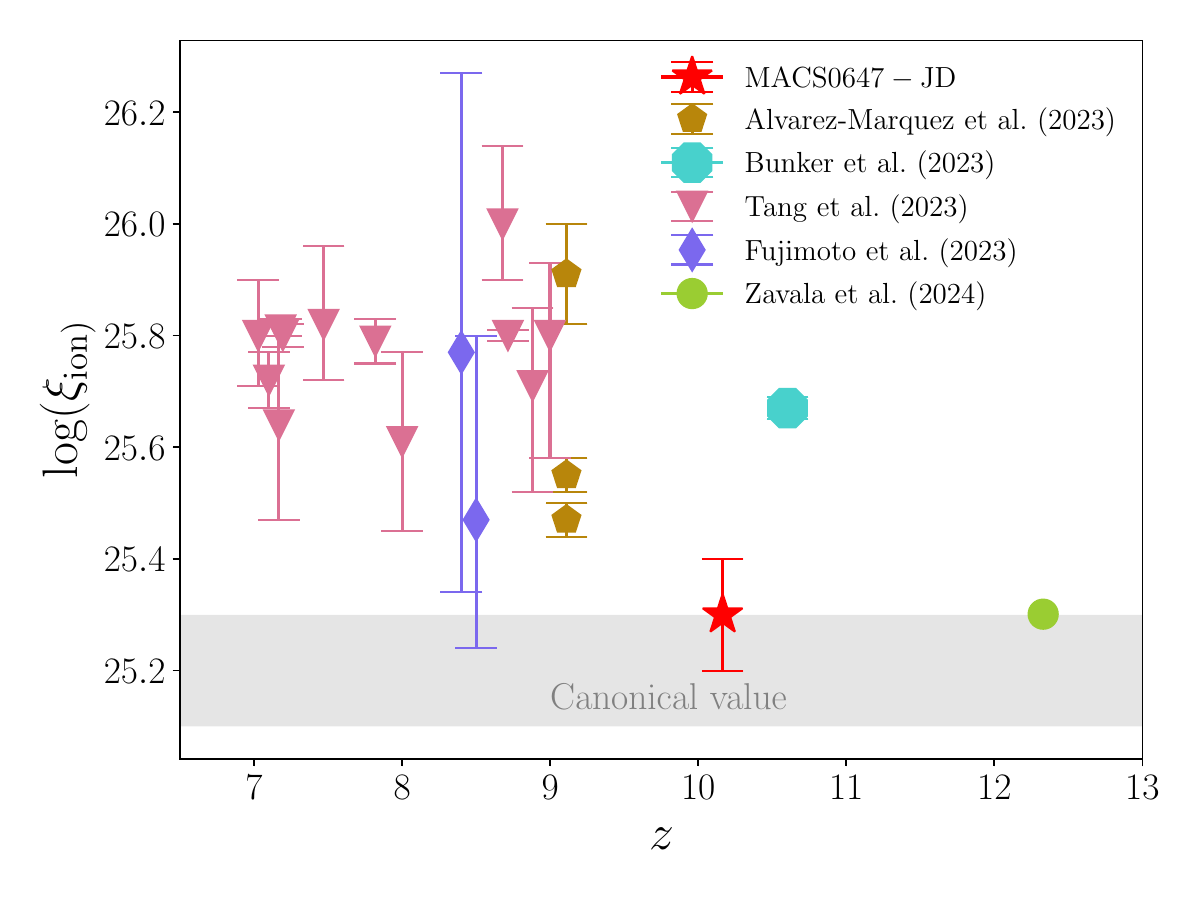}
\caption{The comparison of ${\rm log}(\xi_{\rm ion})$ of MACS0647$-$JD to the previous estimates at $z\gtrsim7$. 
Brown pentagons are MACS1149-JD1 at $z=9.11$ \citep{Alvarez2023}. 
The turquoise hexagon is GNz11 at $z=10.603$ \citep{Bunker2023}. 
Pink triangles represent galaxies at $7<z<9$ \citep{Tang2023}. 
Violet diamonds are median values of NIRCam and HST- selected galaxies at $8<z<9$ \citep{Fujimoto2023}. 
The green point shows GHZ2/GLASS-z12 at $z=12.33$ \citep{Zavala2024}. 
The canonical value of $25.2\pm0.1$ is shown as the gray shaded region \citep{Robertson2013}.
}
\label{fig:xiion_comp}
\end{figure}

The ionizing photon production efficiency can be expressed as 
$\xi_{\rm ion} = \dot{N}_{\rm ion} / L_\nu^{\rm UV}$, 
where $L_\nu^{\rm UV}$ is the UV luminosity and $\dot{N}_{\rm ion}$ stands for the production rate of hydrogen-ionizing photons.
We use H$\alpha$ to derive \Ndotion\ of ($1.2\pm0.1$)\e{54} s\inv,
where ${\rm log}(L_{\rm H\alpha}) = {\rm log} ({\dot{N}_{\rm ion}}) - 11.87$,
assuming zero escape fraction and no dust.
This value should be treated as a lower limit
since a $z\sim10.165$ galaxy likely has some non-zero escape fraction.
Combined with our measured and demagnified UV luminosity
{based on our measured F200W photometry 
(see \S\ref{sec:NIRSpec_NIRCam};} 
\citealt{Hsiao2023b}) for MACS0647$-$JD of $(5.8\pm1.0)\times10^{28}\,{\rm erg\,s^{-1}\,Hz^{-1}}$ \citep[corresponding to $M_{UV} = -20.3 \pm 0.2$; where $M_{UV}=-2.5{\rm log}L^{\rm UV}_{\nu}+51.6$;][]{OkeGunn83},
we estimate an ionizing photon production efficiency 
$\xi_{\rm ion} = \dot{N}_{\rm ion} / L_\nu^{\rm UV} = 2.0\pm0.5$ \e{25} erg\inv\ Hz, or log($\xi_{\rm ion})=25.3\pm0.1$.
This is consistent with the previous estimate log($\xi_{\rm ion})=25.2\pm0.2$ 
based on NIRSpec prism spectroscopy without H$\alpha$ \citep{Hsiao2023b}, and it is similar to the canonical value of 25.2 \citep{Robertson2013}.
It is also similar to the value recently reported for GHZ2/GLASS-z12 at $z = 12.34$
\citep{Zavala2024}.
This aligns with predictions for high-redshift galaxies ($z>8$), especially in younger ($\lesssim10^{8}\,{\rm yr}$) and lower-metallicity ($Z\lesssim Z_{\odot}$) galaxies \citep[e.g.,][]{Schaerer2003,Wilkins2016}.
Other recent JWST studies find somewhat higher
\logxiion\ between 25.4 and 26.0
for high-redshift ($7<z<11$) galaxies \citep[e.g.,][]{Tang2023,Bunker2023,Fujimoto2023,Alvarez2023},
as shown in Figure \ref{fig:xiion_comp}.
Recent studies indicate that fainter galaxies tend to have higher \xiion, 
while brighter galaxies exhibit lower \xiion\ \citep[e.g.,][]{Jung2023,Fujimoto2023,Atek2024}.
{Therefore, it is possible that} our \xiion\ measurement for MACS0647$-$JD falls within the expected range for galaxies of similar UV luminosity ($\rm M_{UV}\sim -20.3$).
{ Similarly bright (or brighter) high-z galaxies
may have similarly high \xiion\ \citep[e.g.,][]{Tang2023,Bunker2023}.
We note those works estimated \xiion\ using SED fitting,
whereas we use H$\alpha$ and UV luminosities in this letter.}

\subsection{Ionization parameter \logU}

The line ratio O32 $=$ \OIIIw\ $/$ \OIIw\ has long been used 
\citep[e.g.,][]{Hicks2002,Papovich2022}
as a diagnostic for the ionization parameter \logU.
We measure a ratio of ${\rm O32}=17\pm2$, which corresponds to log$(U)=-1.9\pm0.1$ adopting the relation from \citet{Papovich2022} 
or log$(U)=-2.0\pm0.1$ adopting the relation from \citet{Diaz2000}.
Note that these relations are not accounting the low-metallcitiy in MACS0647$-$JD.
We also analyze our observed line ratios to estimate the ionization parameter
using 
\HIIC\ version 5.22\footnote{
\href{https://home.iaa.csic.es/~epm/HII-CHI-mistry-opt.html}
{https://home.iaa.csic.es/\about epm/HII-CHI-mistry-opt.html}} \citep{Perez2014}, which yields log$(U)=-1.7\pm0.1$, adopting Binary Population and Stellar Synthesis (BPASS) templates \citep{Stanway2018} and constraints for Star-Forming Galaxies in \HIIC.
All of the log($U$) values derived above are consistent within uncertainties.
The various values in Table \ref{tab:phys} are also consistent within their uncertainties.
These also agree with the previous estimate 
log($U$) $=-1.9\pm0.2$ using Ne3O2 to estimate O32 \citep{Hsiao2023b}
before \OIII\ had been measured in this work.

Our measurement for MACS0647--JD is similar to the value
\logU\ $= -1.8 \pm 0.3$ measured for GHZ2/GLASS-z12 at $z=12.33$
\citep{Castellano2024,Zavala2024}
and \logU\ = $-2.2 \pm 0.9$ measured for GN-z11 at $z = 10.60$ \citep{Bunker2023}.
These and other similar measurements of \logU\ \about\ $-2$ for high-redshift galaxies
reveal strong ionization parameters, comparable to the more starbursting environments we see at low-redshift, but overall higher than the non-star forming galaxies \citep[e.g.,][]{Kewley2002,Mingozzi2024}.





\begin{figure}
\centering
\includegraphics[width=\columnwidth]{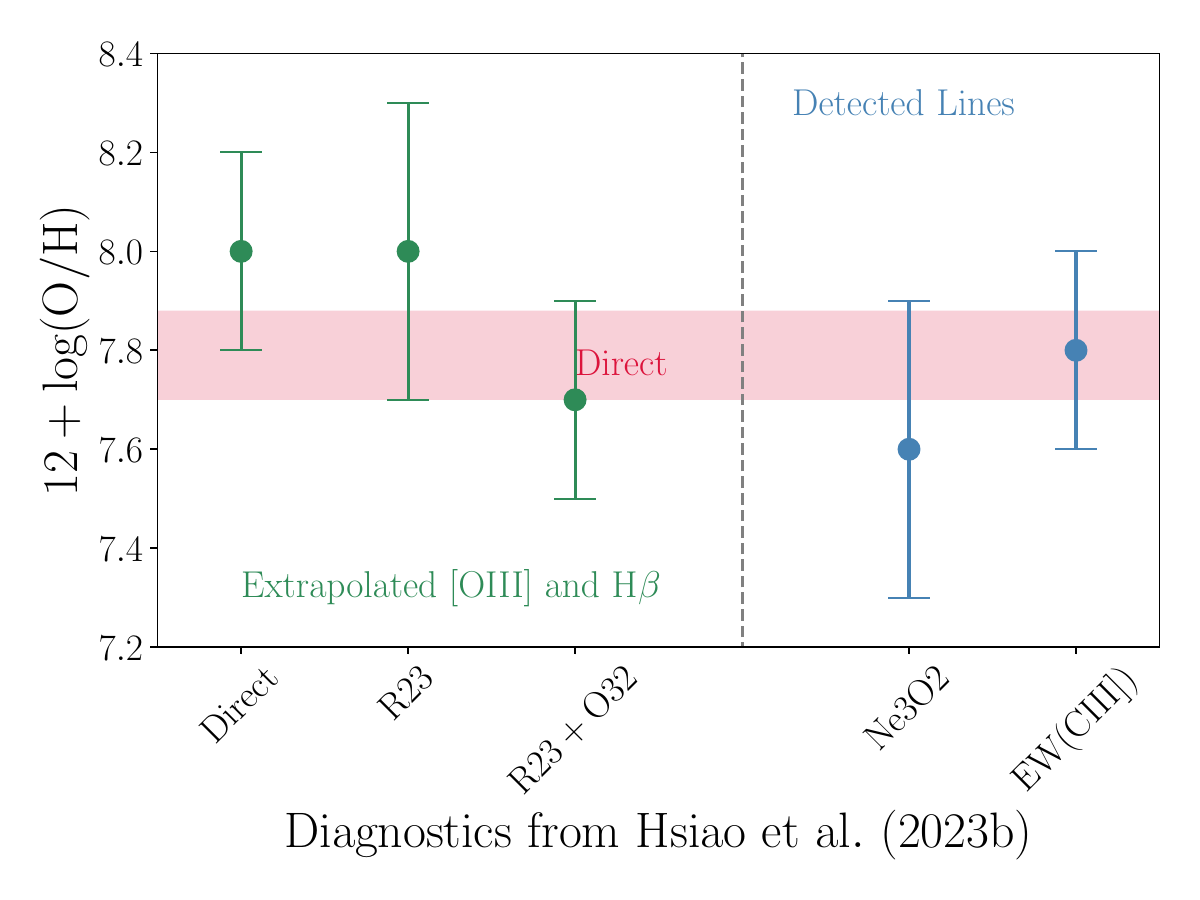}
\caption{MACS0647--JD metallicity measured using the direct-$T_{e}$ method in this paper 
(crimson band)
compared to the previous estimates using different indirect diagnostics in \citet{Hsiao2023b}:
green data points included estimated line fluxes for \OIII\ and \Hbeta,
while the blue data points relied on detected lines only.
Note that for the ``direct" method in the x-axis, \citet{Hsiao2023b} extrapolated \OIII\ and \Hbeta\ from detected lines.
}
\label{fig:metallicity_comp}
\end{figure}

\begin{figure*}
\centering
\includegraphics[width=\textwidth]{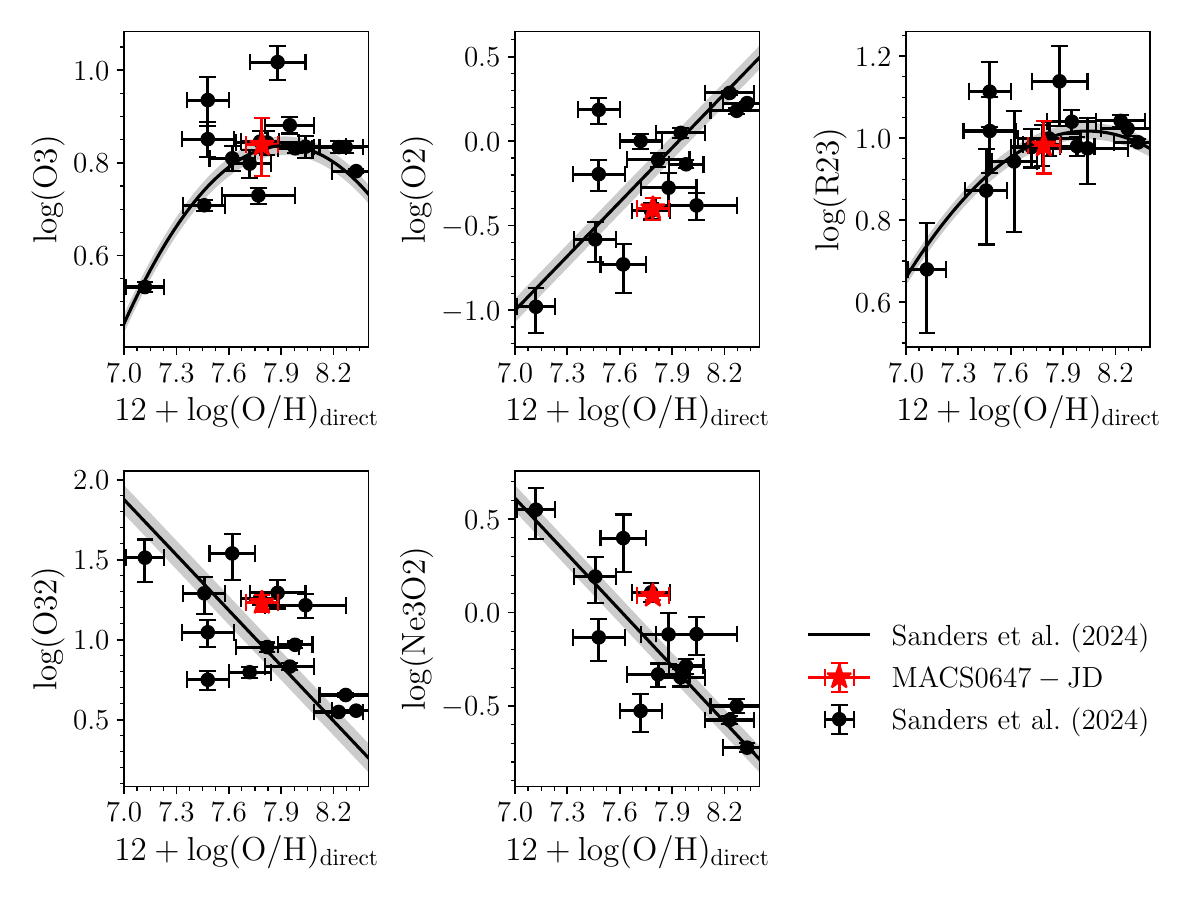}
\caption{\JD\ metallcity using the direct-$T_{e}$ method 
plotted in red against various strong line ratios
measured in \citet{Hsiao2023b}.
Black points show similar measurements for 
galaxies at $z\sim2-9$ from \citet{Sanders2024}.
The line flux ratios are
O3=\OIIIw/\Hbeta; 
O2=\OIIw/\Hbeta; 
R23=(\OIIIdw+\OIIw)/\Hbeta; 
O32=\OIIIw/\OIIw; 
and
Ne3O2=\NeIIIw/\OIIw.
}
\label{fig:diag}
\end{figure*}

\begin{figure}
\centering
\includegraphics[width=\columnwidth]{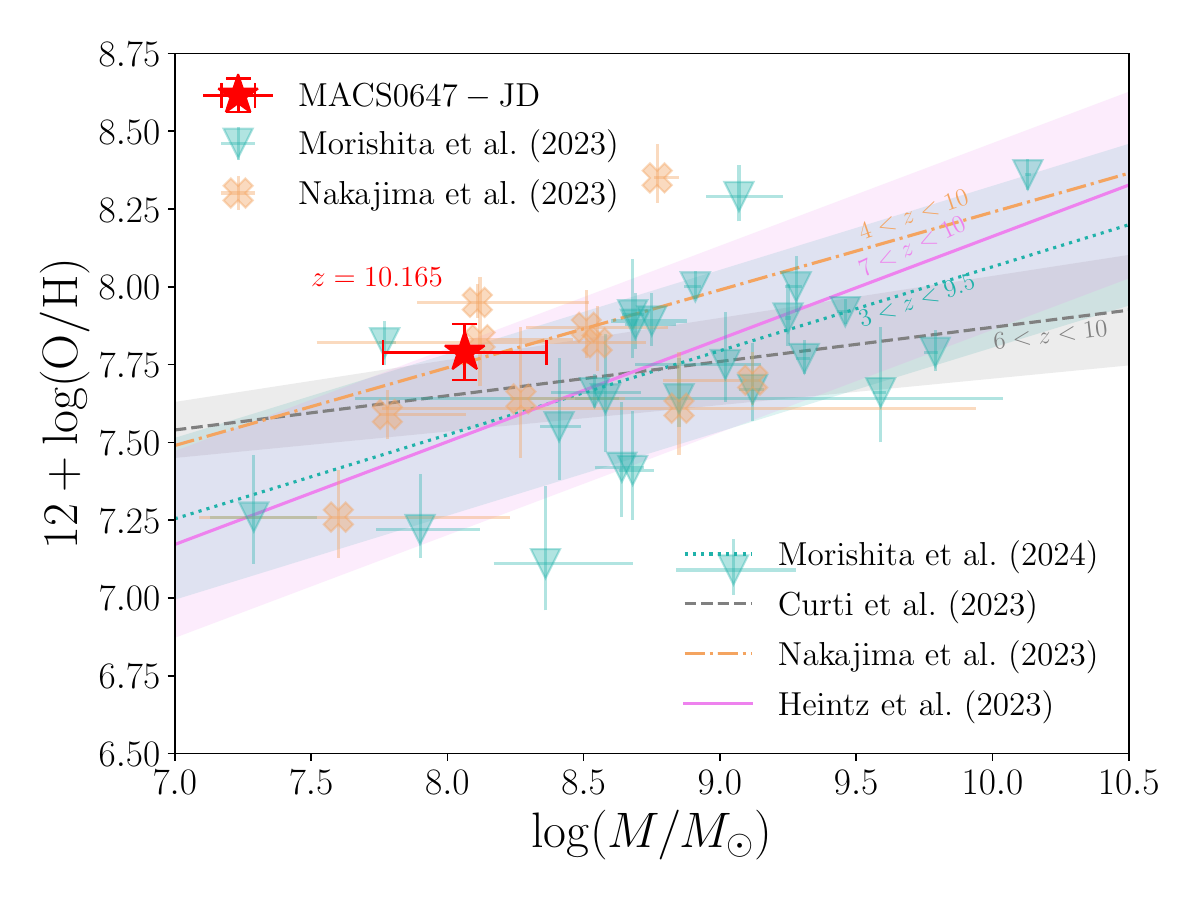}
\includegraphics[width=\columnwidth]{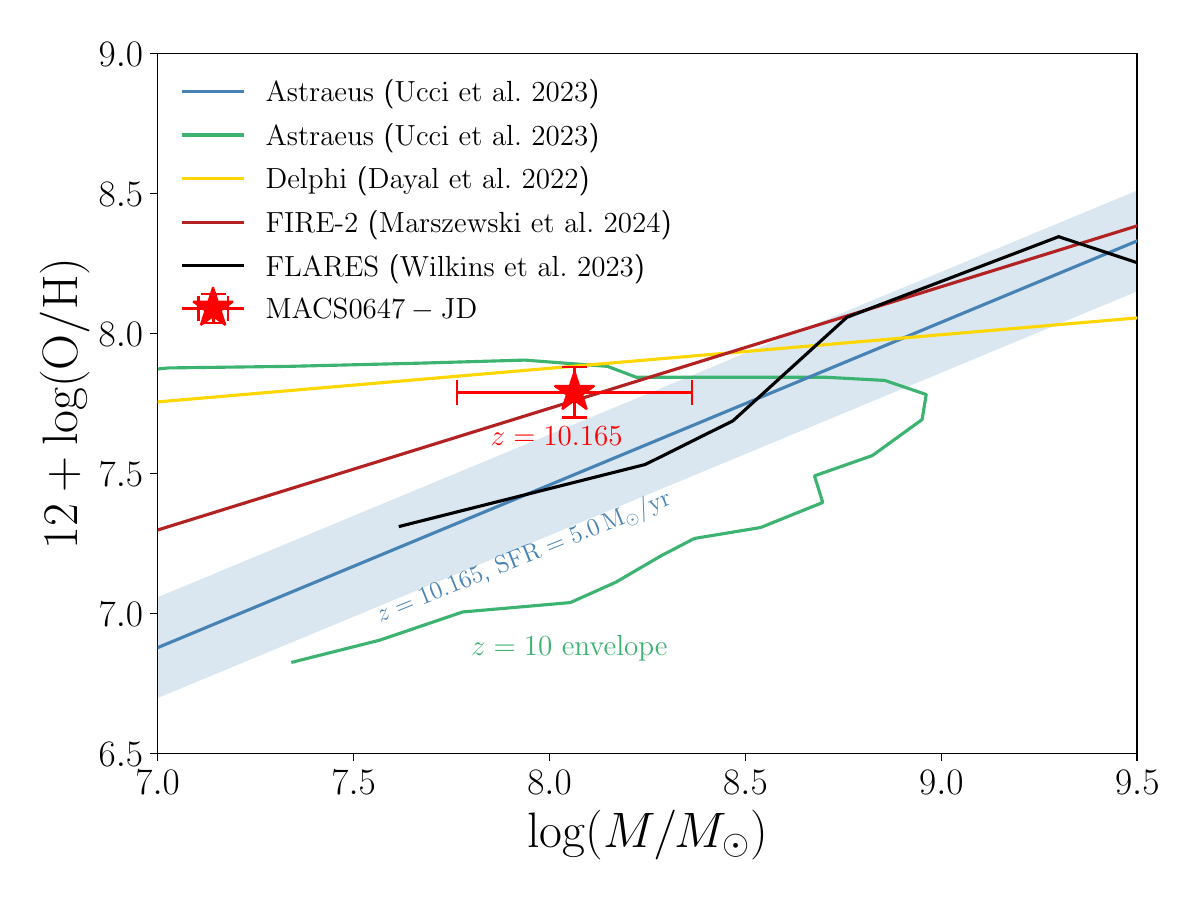}
\caption{
The mass-metallicity relation. MACS0647$-$JD (this work) determined using direct $T_{e}$ method is shown in the red star.
\emph{(Top panel:)} We compare MACS0647$-$JD with some previous works on high redshift galaxies, including 
\citet[][orange Xs and the dot-dashed line]{Nakajima2023} at $4<z<10$, 
\citet[][the violet line]{Heintz2023} at $7<z<10$, 
\citet[][the grey dashed line]{Curti2023} at $6<z<10$, and 
\citet[][tiffany green triangles and the dot line]{Morishita2024} at $3<z<9.5$.
For \cite{Nakajima2023} and \cite{Morishita2024},
we show only galaxies derived from direct $T_{e}$.
Shaded regions represent $1\sigma$ confidence interval of each fit.
\emph{(Bottom panel:)} We also compare MACS0647$-$JD to a number of theoretical predictions \citep{Dayal2022,Ucci2023,Wilkins2023,Marszewski2024}, as marked; the (green) contour shows the extent of the mass-metallicity relation allowed by the ASTRAEUS model \citep{Ucci2023}.
For MACS0647$-$JD, \citet{Heintz2023}, and \citet{Marszewski2024}, we scale the stellar mass by a factor of 0.92 accounting for the initial mass function from \citet{Kroupa2002} to \citet{Chabrier2003} while we scale the stellar mass by a factor of 0.61 for \citet{Dayal2022} and \citet{Ucci2023} from \citet{Salpeter1955} to \citet{Chabrier2003}.
}
\label{fig:mass_metallicity}
\end{figure}

\begin{figure}
\centering
\includegraphics[width=\columnwidth]{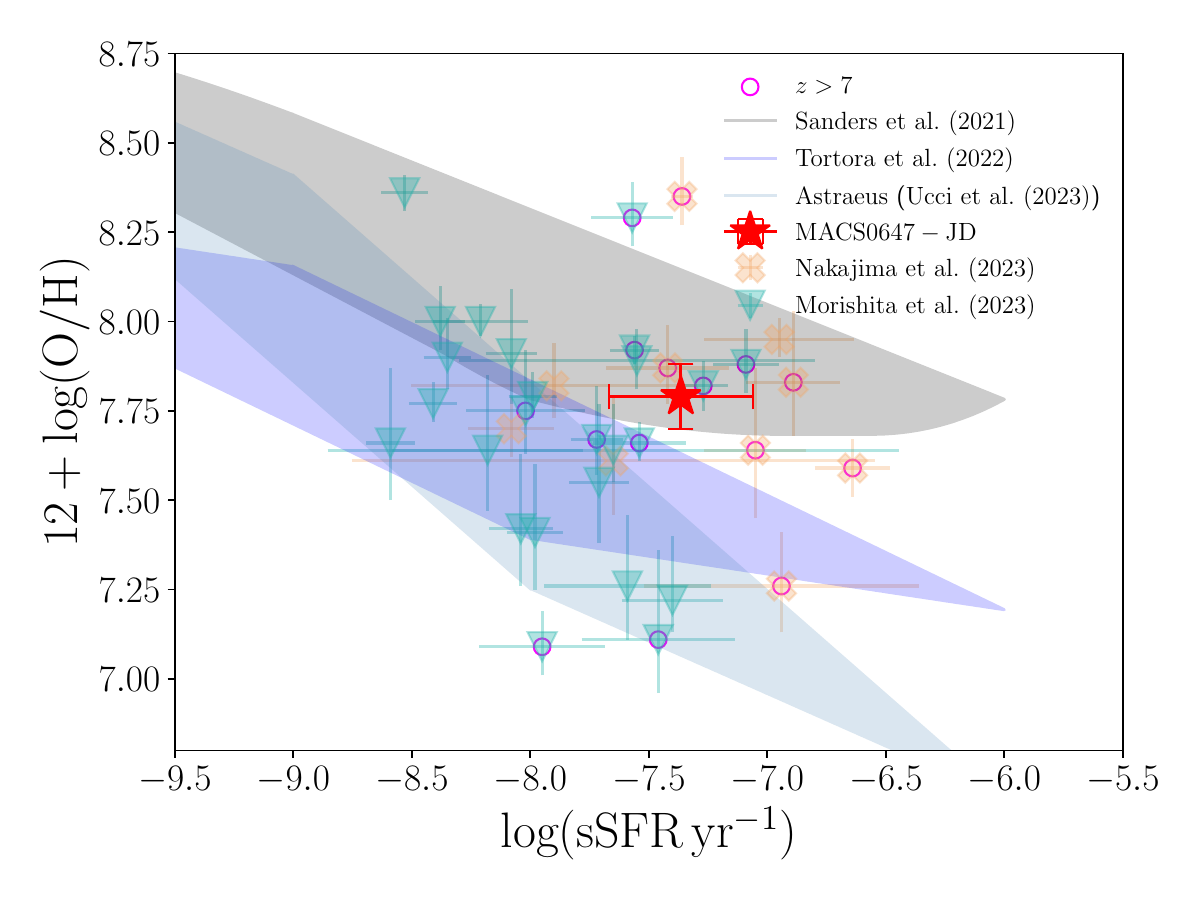}
\caption{Similar to Figure \ref{fig:mass_metallicity} but the fundamental-metallicity relation.
MACS0647$-$JD (this work) determined using direct $T_{e}$ method is shown in the red star.
We compare MACS0647$-$JD with some previous high-redshift direct metallicity measurements
from \citet[][orange Xs]{Nakajima2023} at $4<z<10$ and 
\citet[][tiffany green triangles]{Morishita2024} at $3<z<9.5$.
Galaxies at $z>7$ are specifically marked with magenta circles.
We also compare MACS0647$-$JD to \textsc{Astraeus} \citep{Ucci2023} simulation predictions at $z=10.165$, which is shown using the steelblue shaded region. We also show observationally inferred relations at $z \sim 1.5-3.5$ from \citet{Sanders2021} in the gray shaded region and the relation at $z \sim 0-3.7$ from \citet{Tortora2022} in the purple shaded region.
For MACS0647$-$JD, we scale the stellar mass by a factor of 0.92 and SFR by a factor of 0.94 accounting for the initial mass function from \citet{Kroupa2002} to \citet{Chabrier2003}.
}
\label{fig:fund_metallicity}
\end{figure}

\subsection{Direct Metallicity}
\label{sec:metallicity}

Based on the measured line fluxes of \OIIIw\ from MIRI MRS and the auroral line \OIIIwa\ from NIRSpec, we can obtain a direct metallicity measurement based on electron temperature {with the estimated H$\beta$ as H$\alpha$/2.74 assuming no dust.}
We measure the line flux ratio \OIIIw/\OIIIwa\ $=40\pm5$.
Based on this flux ratio,
we derive the electron temperature of the high-ionization zone $T_{\rm e}$(\OIII) $= 17000 \pm 1000\,$K
using the \pyneb\ \citep{pyneb_Luridiana2015} task \texttt{getTemDen}.
We also find a similar value of $\sim17000\,{\rm K}$ using the equation 4 in \citet{Nicholls2020}.
We then apply the relation
$T_{\rm e}$(\OII) = 0.7 $\times$ $T_{\rm e}$(\OIII) + 3000\,K \citep{Campbell1986} to estimate
the low-ionization zone gas temperature $T_{\rm e}$(\OII) = $15000 \pm 1000\,$K, and the O$^+$/H ionic abundance.
Adding the oxygen abundances from both zones,
we derive a metallicity $12+{\rm log(O/H)}=7.79\pm0.09$,
or $13^{+2}_{-3}\%\,$\Zsun.
We adopt this as our fiducial result.
We also utilize 
\HIIC\ to estimate a metallicity of \logOH\ \about\ $7.56\pm0.12$, 
corresponding to \about $(7^{+3}_{-1})\%$\Zsun.

These results do not vary significantly for density values 
$n_{\rm e}$ between 300 and 3000 cm$^{-3}$,
the range measured for \JD\ in our companion paper 
by resolving the \OII\ doublet \citep{Abdurrouf2024}.

The results also do not depend strongly on our choices
for NIRSpec line fluxes and normalization (\S\ref{sec:normalization}).
Table \ref{tab:phys} presents the results 
with metallicity \logOH\ ranging from 7.7 -- 7.9 (11--16\% \Zsun)
and electron temperature ranging between $15000-19000\,$K.




Our direct metallicity measurement is 
consistent with the range 
$\sim6-20\%\,Z_{\odot}$ estimated by \cite{Hsiao2023b}
using different approaches based on lines including 
\OIIIwa\ and estimates for \OIIIw\ and \Hbeta.
We compare these estimates in Figure \ref{fig:metallicity_comp}.

In Figure \ref{fig:diag}, we plot direct metallicity versus strong line ratios
for \JD\ and galaxies at $z \sim 2$ -- 9 from \cite{Sanders2024}.
Overall, the relation between the direct metallicity and different strong line ratios of MACS0647$-$JD is consistent with galaxies at lower redshift $z \sim 2$ -- 9.




The \JD\ metallicity \logOH\ $= 7.8 \pm 0.1$
is similar to most direct metallicity measurements from JWST for galaxies at $z \sim 4$ -- 9
\citepeg{Nakajima2023,Heintz2023,Curti2023,Morishita2024}.
In Figure \ref{fig:mass_metallicity}, 
we plot metallicity vs.~stellar mass for these galaxies,
including log$(M/M_\odot)=8.1\pm0.3$ measured for \JD\ by \cite{Hsiao2023b} using SED fitting. Given the current limited (and potentially biased) sample of galaxies with direct metallicity measurements, 
we do not detect strong trends with
redshift over the billion years from $z \sim 4$ to 10
or stellar mass from \tentothe{7} to \tentothe{10} \Msun.
If these trends were stronger, we would expect \JD\ to have one of the lowest metallicities
given its relatively low stellar mass and highest redshift among these galaxies.
But we find its metallicity is roughly average compared to the others at $z \sim 4$ -- 10.
This is likely due to a combination of intrinsic scatter and measurement uncertainties.

In the bottom panel of {Figure \ref{fig:mass_metallicity}}, we also show a comparison with theoretical results including those from semi-analytic models \citep{Dayal2022}, semi-numerical models \citep{Ucci2023} and hydrodynamic simulations \citep{Wilkins2023,Marszewski2024}. Although about 0.3 dex higher than the average predicted relation at $z \sim 10$, \JD's metallicity falls within the range predicted by the \textsc{Astraeus} framework \citep{Ucci2023} and is about 0.3 dex higher than that predicted by \textsc{Flares} \citep{Wilkins2023}. Our measurement is in perfect accord with recent results from semi-analytic models tuned against the latest ALMA and JWST data sets \citep[DELPHI;][]{Dayal2022} and the state-of-the-art hydrodynamic simulations \citep[FIRE-2;][]{Marszewski2024}, although these predict extremely different slopes for the mass-metallicity relation. These demarcate the impact of different methodologies and assumptions (metal production, dispersal, the impact of feedback on the gas and metal content, to name a few) on determining the mass-metallicity relation at these early epochs.

In Figure \ref{fig:fund_metallicity}, we plot the Fundamental Metallicity relation where MACS0647$-$JD is again consistent with high redshift ($z \sim 4$ -- 9) galaxies with direct metallicity measurements. 
Given the relatively high specific star formation rate log(sSFR / yr\inv) = $-7.4 \pm 0.3$ for this low-mass system,
the ASTRAEUS model \citep{Ucci2023} predicts 
a lower metallicity \logOH\ $\leq 7.5$ suppressed by outflows.
But the simulation expectations are close given our uncertainties on log(sSFR).
The redshift-dependent extrapolated expectations from \citet{Tortora2022} are similarly close.
Comparing with lower-redshift relations,
we find \JD's metallicity is at the low end of 
$z \sim 1.5-3.5$ observationally-inferred expectations from \cite{Sanders2021}, while more galaxies at $z>10$ with direct metallicity with more precise determinations of SFR and stellar mass are required for more detail discussion.
\section{Conclusions} \label{sec:conclusion}

In this letter, we report MIRI spectroscopic observation of the triply-lensed galaxy merger MACS0647$-$JD at $z=10.165$.
We detect the rest-frame optical emission lines 
\Halpha, \OIIIw, and \OIIIwc\ and measure line fluxes
($23\pm2$) and ($9\pm1$) $\times$ \tentotheminus{18} \cgsfluxunits\ for
\OIIIw\ and \Halpha, respectively, with a line flux ratio $2.5\pm0.4$.
Analyzed jointly with NIRSpec prism line flux measurements presented in \cite{Hsiao2023b}, including the auroral line \OIIIwa,
we present the first direct metallicity measurement of a $z > 10$ galaxy: \logOH\ $= 7.79\pm0.09$, or 0.13$^{+0.02}_{-0.03}\,$\Zsun.
We also measure 
the ionization parameter \logU\ = $-1.9 \pm 0.1$, 
ionizing photon production efficiencty log(\xiion) = $25.3\pm0.1$,
and SFR $5.0\pm0.6\,M_{\odot}/{\rm yr}$ within the past 10 Myr. 

Given the stellar mass log$(M/M_\odot)=8.1\pm0.3$ \citep{Hsiao2023b}
and specific star formation rate log(sSFR / yr\inv) = $-7.4 \pm 0.3$,
we might have expected a lower metallicity \logOH\ \about\ 7.2, 
or \about 3\% \Zsun\ \citep{Ucci2023}.
Instead, the \JD\ metallicity is similar to galaxies at $z \sim 4 - 9$ with direct metallicity measurements \citepeg{Nakajima2023,Curti2023,Morishita2024,Heintz2023}.
With a delensed F200W AB mag 27.2 (\MUV\ = $-20.3$),
\JD\ may be a typical and representative $z \sim 10$ galaxy
fortuitously lensed to be the brightest yet known,
enabling detailed study.

JWST Cycle 3 program GTO 4528 (PI Isaak) 
will observe 13 targets with NIRSpec IFU,
including MACS0647--JD with G395M.
This will be promising to measure total line fluxes,
improve the metallicity precision,
disentangle contributions from components A and B,
and measure any velocity difference to constrain the dynamical mass.
Additionally, they could spectroscopically confirm companion C,
which is not detected in our MIRI MRS data.



MIRI extends the ability of JWST to obtain direct metallicity measurements
for galaxies at $z > 9.5$ with detections of \OIIIw.
What we gain with MIRI is 100 Myr of cosmic time
when direct metallicity measurements are still possible,
pushing the limit from $z \sim 9.5$ (\about500 Myr after the Big Bang)
to $z \sim 11.1$ (\about400 Myr).
Beyond that redshift we will need to swap accurate direct methods for calibrated diagnostics.
Even though these are less precise, this work demonstrates that such diagnostics \textit{can} be relied upon to measure metallicity in high-z systems like MACS0647$-$JD.

MIRI also delivers other important measurements, 
including star formation rate and \xiion, to higher redshifts $z > 11$, as shown with GHZ2/GLASS-z12 at $z=12.33$ \citep{Zavala2024}.

\section{Acknowledgments}
We thank the anonymous referee for useful comments and constructive remarks on the manuscript.

This work is based on observations made with the NASA/ESA/CSA 
\textit{James Webb Space Telescope} (JWST) and \textit{Hubble Space Telescope} (HST). 
The data were obtained from the Mikulski Archive for Space Telescopes (MAST) 
at the Space Telescope Science Institute (STScI), 
which is operated by the Association of Universities for Research in Astronomy (AURA), Inc., 
under NASA contract NAS 5-03127 for JWST. 
We are grateful and indebted to the 20,000 people who worked to make JWST an incredible discovery machine.
These observations are associated with JWST programs GO 4246 and 1433,
and HST GO 9722, 10493, 10793, and 12101.
TH is funded by a grant for JWST-GO-01433 and JWST-GO-04212 provided by STScI under NASA contract NAS5-03127.
TH appreciates the support from the Government scholarship to study abroad (Taiwan).
A is funded by a grant for JWST-GO-01433 and JWST-GO-04246 provided by STScI under NASA contract NAS5-03127.
AA acknowledges support by the Swedish research council Vetenskapsr{\aa}det (2021-05559).
A.B. acknowledges support from the Swedish National Space Administration (SNSA).
EZ acknowledges project grant 2022-03804 from the Swedish Research Council (Vetenskapsr\aa{}det) and has also benefited from a sabbatical at the Swedish Collegium for Advanced Study. A.Z. and L.J.F. acknowledge support by Grant No. 2020750 from the United States-Israel Binational Science Foundation (BSF) and Grant No. 2109066 from the United States National Science Foundation (NSF), and by the Ministry of Science \& Technology, Israel.
J.A.-M., L.C., A.C.-G., C.P.-J. acknowledge support by grant PIB2021-127718NB-100 from the Spanish Ministry of Science and Innovation/State Agency of Research MCIN/AEI/10.13039/501100011033 and by “ERDF A way of making Europe”. C.B.-P. acknowledges support by grant CM21\_CAB\_M2\_01 from the Program ``Garant\'{\i}a Juven\'{\i}l'' from the ``Comunidad de Madrid'' 2021.  PD acknowledge support from the NWO grant 016.VIDI.189.162 (``ODIN") and warmly thanks the European Commission's and University of Groningen's CO-FUND Rosalind Franklin program. 
AZ acknowledges support by Grant No. 2020750 from the United States-Israel Binational Science Foundation (BSF) and Grant No. 2109066 from the United States National Science Foundation (NSF); by the Ministry of Science \& Technology, Israel; and by the Israel Science Foundation Grant No. 864/23.


%

\vspace{5mm}
\facilities{JWST(NIRCam, NIRSpec, MIRI), HST(ACS, WFC3)}

\software{STScI JWST pipeline;
          \msaexp;
          \grizli\ \citep{grizli};
          \astropy\ \citep{astropy2022, astropy2018, astropy2013};
          \piXedfit\ \citep{Abdurrouf2021,Abdurrouf2022};
          \pyneb\ \citep{pyneb_Luridiana2015};
          \HIIC\ \citep{Perez2014};
          \textsc{GALFIT} \citep{Galfit2002}
          }


\appendix


\bibliography{papers}{}
\bibliographystyle{aasjournal}



\end{document}

%% file: newcommands.tex
\newcommand{\LCDM}{$\Lambda$CDM}

\newcommand{\red}[1]{{\color{red} #1}}
\newcommand{\redss}[1]{{\color{red} ** #1}}
\newcommand{\redbf}[1]{{\color{red}\bf #1 \color{black}}}

\newcommand{\ny}{$\tilde {\rm n}$}
\newcommand{\about}{$\sim$}
\newcommand{\appr}{$\approx$}
\newcommand{\gt}{$>$}
\newcommand{\um}{$\mu$m}
\newcommand{\uJy}{$\mu$Jy}
\newcommand{\sig}{$\sigma$}
\newcommand{\Lya}{Lyman-$\alpha$}
\renewcommand{\th}{$^{\rm th}$}
\newcommand{\lam}{$\lambda$}

\newcommand{\tentothe}[1]{$10^{#1}$}
\newcommand{\tentotheminus}[1]{$10^{-#1}$}
\newcommand{\e}[1]{$\times 10^{#1}$}
\newcommand{\en}[1]{$\times 10^{-#1}$}
\newcommand{\cgsfluxunits}{erg$\,$s$^{-1}\,$cm$^{-2}$}
\newcommand{\linefluxunits}{\tentotheminus{20} \cgsfluxunits}

\newcommand{\logU}{$\log(U)$}
\newcommand{\logOH}{12+log(O/H)}

\newcommand{\sinv}{s$^{-1}$}
\newcommand{\kms}{km\,s$^{-1}$}

\newcommand{\footnoteurl}[1]{\footnote{\url{#1}}}

\newcommand{\tnm}[1]{\tablenotemark{#1}}
\newcommand{\super}[1]{$^{\rm #1}$}
\newcommand{\supa}{$^{\rm a}$}
\newcommand{\supb}{$^{\rm b}$}
\newcommand{\supc}{$^{\rm c}$}
\newcommand{\supd}{$^{\rm d}$}
\newcommand{\supe}{$^{\rm e}$}
\newcommand{\supf}{$^{\rm f}$}
\newcommand{\supg}{$^{\rm g}$}
\newcommand{\suph}{$^{\rm h}$}
\newcommand{\supi}{$^{\rm i}$}
\newcommand{\supj}{$^{\rm j}$}
\newcommand{\supk}{$^{\rm k}$}
\newcommand{\supl}{$^{\rm l}$}
\newcommand{\supm}{$^{\rm m}$}
\newcommand{\supn}{$^{\rm n}$}
\newcommand{\supo}{$^{\rm o}$}

\newcommand{\squared}{$^2$}
\newcommand{\cubed}{$^3$}

\newcommand{\sqarcmin}{arcmin\squared}

\newcommand{\supcomma}{$^{\rm ,}$}

\newcommand{\rhalf}{$r_{1/2}$}

\newcommand{\chisq}{$\chi^2$}

\newcommand{\Zgas}{$Z_{\rm gas}$}  
\newcommand{\Zstar}{$Z_*$}  

\newcommand{\per}{$^{-1}$}
\newcommand{\inv}{\per}
\newcommand{\Mstar}{$M^*$}
\newcommand{\Lstar}{$L^*$}
\newcommand{\phistar}{$\phi^*$}

\newcommand{\logM}{log($M_*$/\Msun)}

\newcommand{\LUV}{$L_{UV}$}
\newcommand{\MUV}{$M_{UV}$}

\newcommand{\Msun}{$M_\odot$}
\newcommand{\Lsun}{$L_\odot$}
\newcommand{\Zsun}{$Z_\odot$}

\newcommand{\Mvir}{$M_{vir}$}
\newcommand{\Mt}{$M_{200}$}
\newcommand{\Mf}{$M_{500}$}

\newcommand{\Ndotion}{$\dot{N}_{\rm ion}$}
\newcommand{\xiion}{$\xi_{\rm ion}$}
\newcommand{\logxiion}{log(\xiion)}
\newcommand{\fesc}{$f_{\rm esc}$}

\newcommand{\XHI}{$X_{\rm HI}$}
\newcommand{\XHII}{$X_{\rm HII}$}
\newcommand{\RHII}{$R_{\rm HII}$}

\newcommand{\Halpha}{H$\alpha$}
\newcommand{\Hbeta}{H$\beta$}
\newcommand{\Hgamma}{H$\gamma$}
\newcommand{\Hdelta}{H$\delta$}
\newcommand{\Halphaw}{\Halpha\,$\lambda$6563}
\newcommand{\Hbetaw}{\Hbeta\,$\lambda$4861}
\newcommand{\Hgammaw}{H$\gamma$\,$\lambda$4340}
\newcommand{\Hdeltaw}{H$\delta$\,$\lambda$4101}
\newcommand{\Ha}{\Halpha}
\newcommand{\Hb}{\Hbeta}

\newcommand{\I}{\,{\sc i}}
\newcommand{\II}{\,{\sc ii}}
\newcommand{\III}{\,{\sc iii}}
\newcommand{\IV}{\,{\sc iv}}
\newcommand{\V}{\,{\sc v}}
\newcommand{\VI}{\,{\sc vi}}
\newcommand{\VII}{\,{\sc vii}}
\newcommand{\VIII}{\,{\sc viii}}

\newcommand{\HI}{H\I}
\newcommand{\HII}{H\II}
\newcommand{\HeI}{He\I}
\newcommand{\HeII}{He\II}

\newcommand{\CII}{[C\II]}
\newcommand{\CIIw}{\CII\,$\lambda$2325 (blend)}
\newcommand{\CIII}{[C\III]}
\newcommand{\CIIIw}{\CIII\,$\lambda$1908}
\newcommand{\CIIId}{C\III]}
\newcommand{\CIIIdw}{C\III]\,$\lambda\lambda$1907,1909}
\newcommand{\CIV}{C\IV}
\newcommand{\CIVw}{\CIV\,$\lambda$1549}
\newcommand{\OII}{[O\II]}
\newcommand{\OIIw}{\OII\,$\lambda$3727}
\newcommand{\OIIdw}{\OII\,$\lambda\lambda$3727,3729}
\newcommand{\OIII}{[O\III]}
\newcommand{\OIIIw}{\OIII\,$\lambda$5008}
\newcommand{\OIIIww}{\OIII\,$\lambda$4960,$\lambda$5008}
\newcommand{\OIIIdw}{\OIIIww}
\newcommand{\OIIIwa}{\OIII\,$\lambda$4363}
\newcommand{\OIIIwc}{\OIII\,$\lambda$4960}
\newcommand{\NeIII}{[Ne\III]}
\newcommand{\NeIIIw}{\NeIII\,$\lambda$3869}
\newcommand{\NeIIIwb}{\NeIII\,$\lambda$3968}
\newcommand{\NII}{[N\II]}
\newcommand{\NIIw}{\NII\,$\lambda$6585}
\newcommand{\NIIww}{\NII\,$\lambda$6550,$\lambda$6585}
\newcommand{\SII}{[S\II]}
\newcommand{\SIIww}{\SII\,$\lambda$6718,$\lambda$6733}
\newcommand{\HeIw}{HeI\,$\lambda$3889}
\newcommand{\HeIwa}{HeI\,$\lambda$4473}
\newcommand{\HeIIw}{HeII\,$\lambda$1640}
\newcommand{\NIII}{N\III]}
\newcommand{\NIV}{N\IV]}
\newcommand{\NIIIw}{\NIII\,$\lambda$1748}
\newcommand{\NIVw}{\NIV\,$\lambda$1486}
\newcommand{\MgII}{Mg\II}
\newcommand{\MgIIw}{\MgII\,$\lambda$2800}

\newcommand{\Lyaw}{Ly$\alpha$\,$\lambda$1216}



\newcommand{\Om}{\Omega_{\rm M}}
\newcommand{\OL}{\Omega_\Lambda}

\newcommand{\etal}{et al.}

\newcommand{\citeps}{\citep}

\newcommand{\HST}{{\em HST}}
\newcommand{\SST}{{\em SST}}
\newcommand{\Hubble}{{\em Hubble}}
\newcommand{\Spitzer}{{\em Spitzer}}
\newcommand{\Chandra}{{\em Chandra}}
\newcommand{\JWST}{{\em JWST}}
\newcommand{\Planck}{{\em Planck}}

\newcommand{\Bradac}{{Brada\v{c}}}

\newcommand{\citepeg}[1]{\citep[e.g.,][]{#1}}

\newcommand{\range}[2]{\! \left[ _{#1} ^{#2} \right] \!}  

\newcommand{\grizli}{\textsc{grizli}}
\newcommand{\eazypy}{\textsc{eazypy}}
\newcommand{\msaexp}{\textsc{msaexp}}
\newcommand{\trilogy}{\textsc{trilogy}}
\newcommand{\bagpipes}{\textsc{bagpipes}}
\newcommand{\beagle}{\textsc{beagle}}
\newcommand{\photutils}{\textsc{photutils}}
\newcommand{\SEP}{\textsc{sep}}
\newcommand{\piXedfit}{\textsc{piXedfit}}
\newcommand{\pyneb}{\textsc{pyneb}}
\newcommand{\HIIC}{\textsc{hii-chi-mistry}}
\newcommand{\astropy}{\textsc{astropy}}
\newcommand{\astrodrizzle}{\textsc{astrodrizzle}}
\newcommand{\multinest}{\textsc{multinest}}
\newcommand{\cloudy}{\textsc{Cloudy}}
\newcommand{\jdaviz}{\textsc{Jdaviz}}

\renewcommand{\tt}[1]{\texttt{#1}}

\newcommand{\SE}{\tt{SourceExtractor}}

\newcommand{\PD}[1]{\textcolor{blue}{[PD: #1\;]}}

\newcommand{\JD}{MACS0647$-$JD}

%% file: lines.tex
\begin{deluxetable*}{lcccccc}
\tablecaption{\label{tab:lines}MIRI MRS measured emission line fluxes for \JD1.
}
\tablewidth{\columnwidth}
\tablehead{
\colhead{Emission} &
\colhead{Rest} &
\colhead{Observed} &
\colhead{Redshift} &
\colhead{Observed} &
\colhead{Observed} &
\colhead{Intrinsic} 
\\[-0.2cm]
\colhead{Line} &
\colhead{wavelength} &
\colhead{wavelength} &
\colhead{$z$} &
\colhead{Flux\supa} &
\colhead{FWHM} &
\colhead{FWHM}
\\[-0.2cm]
\colhead{} &
\colhead{$\rm \AA$}  &
\colhead{$\rm \mu m$} &
\colhead{} &
\colhead{$10^{-19}\,$erg/s/cm$^2$} &
\colhead{\kms} &
\colhead{\kms}
%
}
\startdata
\OIII 
& 4960.30
& 5.5379$^{b}$
&
& $75$$^{c}$%
& 
& 
\\
\OIII 
& 5008.22 
& $5.5915 \pm 0.0002$
& $10.1644 \pm 0.0004$
& $226\pm21$
& $206\pm21$
& $191\pm19$
\\
\Halpha\
& 6564.62
& $7.3300 \pm 0.0002$
& $10.1659 \pm 0.0003$
& $90\pm10$
& $150\pm19$
& $127\pm16$
\enddata
\tablenotetext{a}{JD1 line fluxes are not corrected for lensing magnification ($\mu=8$).}
\tablenotetext{b}{\OIIIwc\ wavelength, redshift, and FWHM fixed to those of \OIIIw.}
\tablenotetext{c}{\OIIIwc\ flux assigned \OIIIw\ flux divided by 2.98.}
\end{deluxetable*}

%% file: properties2.tex
\begin{deluxetable*}{ccccc}
\tablecaption{\label{tab:phys}Emission line ratios and physical properties estimated for MACS0647$-$JD.}
\tablewidth{\columnwidth}
\tablehead{
\colhead{} &
\multicolumn{2}{c}{NIRSpec prism} &
\multicolumn{2}{c}{Normalized to H$\alpha$/H$\gamma$=6.11} \\[-0.2cm]
\colhead{} &
\colhead{Stack$^{a}$} &
\colhead{Obs 21 JD2} &
\colhead{Stack} &
\colhead{Obs 21 JD2}}
\startdata
\multicolumn{5}{c}{Emission Line Ratios} \\ \hline
\OIIIw\ $/$ $\lambda$4363 & $40\pm5$ & $48\pm16$  & $44\pm5$ & $78\pm24$  \\
O32 = \OIIIw\ $/$ \OIIw\ & $17\pm2$& $13\pm2$ & $19\pm1$ & $20\pm1$ \\
H$\alpha$ / H$\gamma$ & $5.5\pm0.7$  & $3.8\pm0.6$ &  \multicolumn{2}{c}{6.11 (fixed)}\\
R3 = \OIIIw\ $/$ \Hbeta$^{b}$ & \multicolumn{4}{c}{$6.9\pm1.0$}\ \\
\hline \hline   
\multicolumn{5}{c}{Physical Properties} \\ \hline  
12+log(O/H)$^{c}$ & $7.79\pm0.09$ & $7.90\pm0.15$ & $7.84\pm0.09$ & $8.10\pm0.14$ \\ 
$Z / Z_\odot$  &  
$0.13^{+0.03}_{-0.02}$ & 
$0.16^{+0.07}_{-0.05}$ & 
$0.14\pm0.03$ & 
$0.26^{+0.09}_{-0.07}$ \\
log($U$) & $-1.9\pm0.1$ &$-2.0\pm0.1$&$-1.9\pm0.1$& $-1.7\pm0.1$\\
SFR ($M_{\odot}\,{\rm yr^{-1}}$) &  \multicolumn{4}{c}{$5.0\pm0.6$}\\
log(\xiion) & \multicolumn{4}{c}{$25.3\pm0.1$}\\
$T_{e}(\rm{[OIII]})$(K) & $17000\pm1000$ & $15000\pm2000$ & $16000\pm1000$ & $13000\pm1000$ 
\enddata
\tablenotetext{a}{We adopt this configuration as fiducial values throughout the letter.}
\tablenotetext{b}{\Hbeta\ is estimated as \Halpha\ $/$ 2.74, {\bf assuming no dust}.}
\tablenotetext{c}{Estimated using \pyneb. Other methods gave similar results
(see \S\ref{sec:metallicity}).}
\end{deluxetable*}